\documentclass[fleqn,usenatbib]{mnras}

\usepackage{newtxtext,newtxmath}

\usepackage[T1]{fontenc}

\DeclareRobustCommand{\VAN}[3]{#2}
\let\VANthebibliography\thebibliography
\def\thebibliography{\DeclareRobustCommand{\VAN}[3]{##3}\VANthebibliography}

\usepackage{graphicx}	
\usepackage{amsmath}	

\usepackage{xcolor}
\usepackage{newtxtext,newtxmath}

\newcommand{\som}{\,$\rm{M_{\odot}}$}

\title[Post-Impact Planetesimals Sustaining EDDs]{Post-Giant Impact Planetesimals Sustaining Extreme Debris Disks}

\author[L. Watt et al.]{
Lewis Watt$^{1}$,
Zo\"e M. Leinhardt$^{1}$,
Philip J. Carter$^{1}$\thanks{E-mail: p.carter@bristol.ac.uk}
\\
$^{1}$School of Physics, University of Bristol, H.H. Wills Physics Laboratory, Tyndall Avenue, Bristol BS8 1TL, UK\\
}

\date{Accepted XXX. Received YYY; in original form ZZZ}

\pubyear{2022}

\begin{document}
\label{firstpage}
\pagerange{\pageref{firstpage}--\pageref{lastpage}}
\maketitle

\begin{abstract}
Extreme debris disks can show short term behaviour through the evolution and clearing of small grains produced in giant impacts, and potentially a longer period of variability caused by a planetesimal population formed from giant impact ejecta. In this paper, we present results of numerical simulations to explain how a planetesimal populated disk can supply an observed extreme debris disk with small grains. We simulated a sample of giant impacts from which we form a planetesimal population. We then use the $N$-body code {\sc Rebound} to evolve the planetesimals spatially and collisionally. We adopt a simplistic collision criteria in which we define destructive collisions to be between planetesimals with a mutual impact velocity that exceeds two times the catastrophic disruption threshold, $V^*$. We find that for some configurations, a planetesimal populated disk can produce a substantial amount of dust to sustain an observable disk. The semi-major axis at which the giant impact occurs changes the mass added to the observed disk substantially while the orientation of the impact has less of an effect. 
We determine how the collision rate at the collision point changes over time and show that changes in semi-major axis and orientation only change the initial collision rate of the disk. Collision rates across all disks evolve at a similar rate.
\end{abstract}




\section{Introduction}


The last stage of planetary growth occurs after most of the gas from a protoplanetary disk has dissipated -- what remains is the solid debris, ranging from large planetary embryos to small dust grains.
 Without the gas to damp the gravitational excitations between large planetary embryos, eccentricity in the disk grows and orbital crossings become more common. 
Hence, the end stage of terrestrial planet formation begins and is dominated by giant impacts between the planetary embryo seeds \citep{Chambers1998-Gasremoved-cols,Agnor_1999_Large_impacts,Morishima_2010_planetesimals_to_terrestrial_planets,Elser_2011_How_common_Earth_Moon_systems}.

Evidence of the giant impact stage of planetary growth is located close to home in our very own solar-system. The Earth-Moon system is almost certainly formed from at least one giant impact between the proto-Earth and another planetary embryo \citep{Hartmann_&_Davis_1975_Lunar_origin,Canup2004-LunarFormation,Cuk_&_Stewart_2012_Moon_fast_spinning_Earth,Canup_2012_Forming_Moon_Earth_like_comp,rufu+2017_multi_col_earth_moon,Lock+2018_moon_origin}. 
The high core-to-mantle ratio of Mercury has been proposed to be caused by a giant impact \citep{Benz-2007-origin-of-mercury} and outside our solar system there is indication that some super-Earths show signs of giant impacts having an effect on their formation \citep{Bonomo2019_Kepler107impact,Denman2020}. As giant impacts happen on extremely short-timescales relative to the age of the systems they occur in, they would be extremely difficult to directly observe even if it was possible with current observing technology. However, giant impacts are a violent affair producing substantial ejecta. A proportion of the ejecta that forms small grains can be observed as an infrared excess in the spectral energy distribution of the star. 
The ejecta that escapes the giant impact forms a disk \citep{Jackson2012_Moon,Jackson-2014-planetary-collisions-at-large-au}. 

Observed debris disks are an indicator of successful planet formation. Debris disks are analogues to our Kuiper belt, with the mass of the disk mainly residing in large planetesimals and a collisional cascade producing the small dust which makes up the surface area which we observe. 
Within the last decade, a handful of very bright debris disk systems have been observed classed as Extreme Debris Disks (EDDs).  
EDDs have distinctive characteristics that separate them from traditional debris disks, such as having fractional luminosities, $f=L_d/L_*$ (where $L_d$ is the disk luminosity and $L_*$ is the stellar luminosity), exceeding $10^{-2}$; higher temperatures; and showing variability on much shorter timescales of years instead of the millions of years typical of a traditional debris disk. ID8 \citep{Meng2014-ID8science,Meng2015-pc,Su-2019-extreme-disk-variability} is an example of a system with yearly and sub-yearly variation with periodicity, TYC 8241 2652 1 was a bright disk with consistent mid-IR luminosity between 2008 and 2010 before decreasing rapidly by at least a factor of 30 \citep{Melis_2012,2017A&A...598A..82G}, and HD 23514 shows warm excess that suggests the disk sits close to the host star \citep{rhee_2008}. 
Traditional debris disks, for comparison, often have a fractional luminosity of $f<10^{-4}$ and typically are observed at longer wavelengths suggesting a colder disk and hence sitting at 10s of au away from the star. An example outside our Solar System is $\beta-$Pictoris which was the first debris disk to be spatially observed \citep{1984Sci...226.1421S,1998Natur.392..788H}. 

A typical debris disk has a slow collisional evolution, on the scale of hundreds of millions of years, in which fine, micron sized dust is produced. The micron-sized dust produces an excess infrared emission when irradiated by the central star. The dust is lost over time due to processes like Poynting-Robertson drag and radiation pressure blowout. The dust is replenished through a steady-state collisional cascade, where the largest objects in the disk are ground down over time \citep{Wyatt-2008-evolution-of-debris-disks}. These largest objects are known as planetesimals and make up the bulk of the mass in a disk. The lifetime of the disk is then set by the collisional activity between the boulders.  

The evolution of the lightcurves of EDDs does not follow what we would expect of a tradional debris disk which follows a steady state collsional cascade on a timescale of millions of years. Many EDDs have been observed to include quiescent and active states, where the active state is much brighter, indicating a sudden increase in the amount of debris in the system, and shows variability on shorter timescales than traditional debris disks  \citep{rhee_2008,Melis_2010,Melis_2012,Meng2012-variability}. Some EDD systems, such as ID8, have been known to have multiple stages of active and quiescent states, indicating a varying level of debris in the system  \citep{Meng2014-ID8science,Su-2019-extreme-disk-variability}. 
A likely source of this excess debris is giant impacts.

Most giant impacts produce a considerable amount escaping material (most giant impacts are not perfect merging events; \citealt{Leinhardt-2012-collisions-between-gravity-dom-bodies,2017E&PSL.470...87G}). It is common in these large impacts for a significant amount of the escaping mass to be vaporised \citep{2020JGRE..12506042C,Gabriel_2021_vapour_impacts,Watt_2021_vapour_EDD}. Vaporised mass is favoured to explain EDD systems as vapour can quickly condense into small fragments/spherules ranging from microns to centimetres in size \citep{Johnson2012-vaporplumes}. Vapour condensate can explain how a disk can go from a quiescent state to an active state extremely fast as observable material is formed almost instantaneously after a collision. If the make up of post-impact debris is mostly small grains initially, it can also explain how a disk can then quickly transition back into a quiescent state as most of the condensate will be lost through radiation processes by the star on short-timescales \citep{Su-2019-extreme-disk-variability}. Many EDDs exist around stars with ages within the terrestrial planet formation age, 10-200 Myr \citep{Meng2015-pc,2017ApJ...836...34M,Moor_2021_warm_DD_sample}. The overlap in timescale supports the giant impact explanation as the main cause for EDDs. However, it should be noted that a few EDDs exist around older stars  \citep{Melis_2021_EDD_intermediate_age}. Explanations for such systems include a potential late stage instability, causing late time giant impacts. Another possible explanation is a wide binary causing instabilities in the system \citep{Zuckerman_2015_wide_companions,Moor_2021_warm_DD_sample}.  

If the initial EDD is observable mostly through small grains formed from vapour, the question then becomes how are these disks sustained? Current debris disk models suggest that a disk close to the star is made of just small grains would dissipate on timescales shorter than their observed lifetimes \citep{Wyatt-2008-evolution-of-debris-disks}. Current models do not support shielding of smaller than blowout sized grains in optically thick clumps, which would increase the lifetime of the disk even when the geometry of giant impact produced debris results in an increase in collisional activity \citep{Jackson-2014-planetary-collisions-at-large-au}. Another hypothesis is that two debris populations are formed by a giant impact: one population being mostly small grains formed from vapour condensate, and the other population being larger planetesimal objects formed from the escaping melt material. If the melt material formed a significant background population of planetesimals, these objects can then form the top end of a collisional cascade which can supply the disk with small grains over time. ID8 would suit such a scenario, as after multiple peaks and dips in its light curve, after 2017 there is a steady rise in excess flux suggesting a gradual release of small grains into the system \citep{Su-2019-extreme-disk-variability}.

The goal of this paper is to further understand the two population model of small grain production after a giant impact has occurred. Our focus will be on the planetesimal population formed post-impact, and its collisional evolution in the stellar system. In section \ref{sec:methods} we outline how the simulations were set up for giant impacts using a smoothed particle hydrodynamical (SPH) code, and the $N-$body set up for spatially and collisionally evolving the disk post-giant impact. In section \ref{sec:results} we present the results of our simulations through exploring the parameter space to see how the evolution of the planetesimal population varies. Section \ref{sec:discussion} discusses how our results affect the observability of an EDD as well as the limitations of our study. We summarise our conclusions in section \ref{sec:conclusion}.

\section{Methods}
\label{sec:methods}
Below we summarise the SPH simulations we conducted, how fragments post-collision are determined, and how the $N-$body simulations were set up from the planetesimal population derived from the SPH simulations.

\subsection{Gadget-2 Simulations}

Extreme debris disks can be formed from the debris of giant impacts. Since extreme debris disks are usually located around the terrestrial region of young stars \citep{Moor_2021_warm_DD_sample}, we focus on simulating impacts between rocky embryos. To do so we use a modified version of Gadget-2 \citep{Springel-2005-Gadget-2,Marcus2009-ANEOS,Cuk_&_Stewart_2012_Moon_fast_spinning_Earth,phil_carter_2022_gadget2} which allows us to use tabulated equations of state. We use the ANEOS equations of state \citep{Melosh2007-ANEOS,Marcus2009-ANEOS,old_EOS} for iron and forsterite  to model the core and mantle respectively. The model embryos have a core to mantle mass ratio of 3:7 and are given initial temperature profiles from \citet{Valencia-2005-internal-structure}. Each embryo is initialized and equilibrated similarly to previous work \citep[e.g.][]{Carter2018,Denman2020,Watt_2021_vapour_EDD}. Each embryo was equilibrated using velocity damping and locked to a set specific entropy for each material and then each embryo was allowed to equilibrate without any restrictions. The values for each embryo are given in Table \ref{tab:gadget2setup}.

\begin{figure*}
    \centering
    \includegraphics[width=\linewidth]{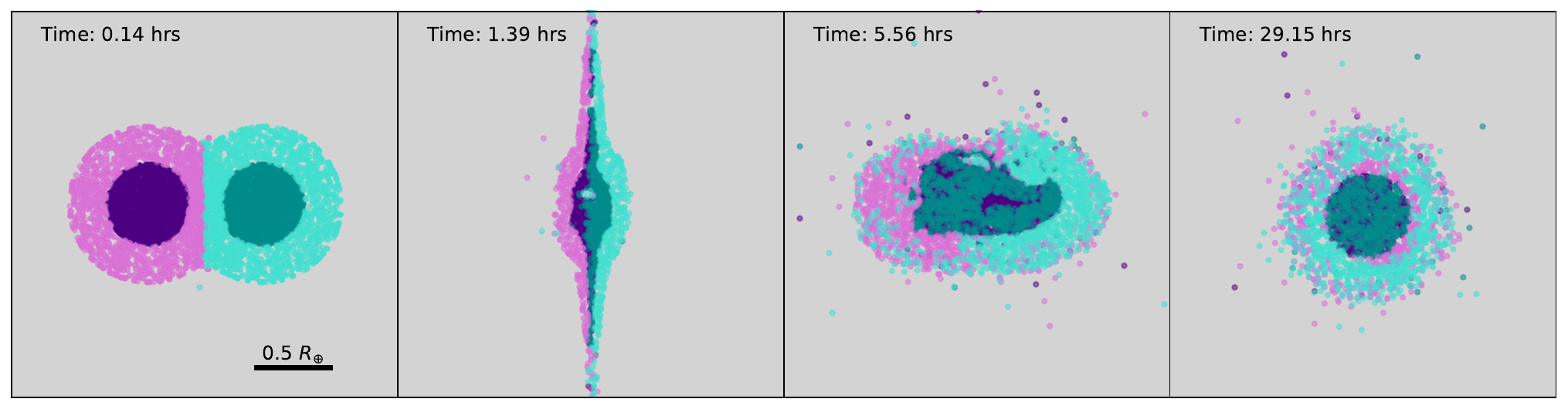}
    \caption{Four snapshots of giant impact 1 from Table \ref{tab:giant_impacts} with a cut in the z-direction to observe a midplane slice. Time for each snapshot is given in the top left corner of each subplot. Core material is shown in darker colours while mantle material is shown in lighter colours. The core and mantle are iron and forsterite respectively.}
    \label{fig:giant_impact_example}
\end{figure*}
In this study, we need to know to a reasonable accuracy the number of planetesimals formed post-impact in order to ascertain the mass in the planetesimal disk.
Within the SPH simulations, we find a number of gravitationally bound groupings of escaping particles. The number of groupings will depend on the particle number resolution. We assume these groupings will form planetesimal sized objects. These planetesimals will form the top end of a mass distribution of a planetesimal population. The number of smaller planetesimals is estimated from the combined mass of the non-grouped escaping particles.
In order to resolve a significant number of planetesimals forming in our chosen simulations, we simulate two giant impacts with a total particle number of $4\times 10^5$ and one giant impact with a total particle number of $4\times 10^4$. We chose three unique giant impacts in order to understand how different impacts effect the outcome of extreme debris disks. The simulated impacts are listed in Table \ref{tab:giant_impacts} with the collision parameters and outcomes. Fig. \ref{fig:giant_impact_example} shows the evolution of giant impact 1. From these impacts we determine a population of escaping planetesimals that will feed the observed disk over time. 

\subsection{Determining Planetesimal Distribution}
\label{sect:methods_plan_distrib}

The focus of this paper is to understand the effect that planetesimals formed from giant-impacts have on the evolution of an extreme debris disk. In order to do so, we first must determine the planetesimals/fragments that form from the post-impact debris. The largest planetesimals will be resolved as bound clumps of particles in the SPH simulations. We define these as `grouped' planetesimals. The grouped planetesimals will not make up the total mass of all planetesimals formed from the giant impact. Planetesimals will form from melt mass found in unbound SPH particles (particles not found in gravitationally bound clumps). We call these `non-grouped' planetesimals. The full planetesimal size distribution is then a mixture of grouped planetsimals which make up the largest planetesimals and non-grouped planetesimals that make up the smaller planetesimals.

 Using the method outlined in section 2.2.1 of \citet{Watt_2021_vapour_EDD} we recursively find gravitationally bound groupings of particles down to a minimum of 5 particles. This gives us a list of large remnants found post-giant impact which will also include planetary embryo(s) if the collision was not super-catastrophic. We remove the largest remnant and sometimes the second largest remnant if the mass ratio between the largest and second largest remnants is greater than 0.2. 
The number of planetesimals varies depending on the snapshot we choose post-collision for the simulated giant impacts, so we also apply a cut to the grouped planetesimals. Any bound group of fewer than 100 particles we choose to remove, as this gives a stable number of planetesimals regardless of snapshot choice while still giving a large number of resolved planetesimals. Therefore, any bound group with greater than 100 particles that is not the largest (or second largest) remnant are labelled as grouped planetesimals. The total mass of the bound groupings with less than 100 particles is added to the unbound escaping mass.  

The SPH giant impact simulations can resolve the largest escaping planetesimals, but this will only make up a small amount of the total escaping planetesimal population post-impact. The grouped planetesimals will have a minimum mass associated with the 100 particle lower limit we set. The minimum planetesimal mass resolved for a given giant impact will then depend on the resolution of the impact. In order to simulate a collisionally active disk of planetesimals, we need to create a distribution of smaller planetesimals from the non-grouped escaping particles to add to the distribution of grouped planetesimals, which we label as non-grouped planetesimals. First we determined the non-grouped mass that will form planetesimals. For all escaping unbound particles, we calculate the percentage of vapour in each particle. The vapour fraction of an SPH particle is determined using the lever rule at the triple-point temperature. We use 1890$^{\circ}$C and $2970^{\circ}$C for forsterite \citep{Nagahara1994-forsteritetp} and iron \citep{Liu1975-TPIron} respectively.  For any escaping particle with a vapour fraction less than or equal to 10 per cent we add the mass to the total mass which forms non-grouped planetesimals. Any particle with a vapour fraction above 10 per cent we assume will form into smaller sized objects as the expanding vapour will limit the gravitational in-fall and reaccumulation of material and therefore ignore in this study. 

After the total mass of non-grouped planetesimals has been defined, a size distribution must be set in order to generate a population of planetesimals. Debris disks typically have a size distribution of $n(D) \propto D^q$ \citep{Wyatt-2008-evolution-of-debris-disks}, with $q$ being determined to have a value between $-3$ and $-4$. A larger value of $q$ will lead to the total mass in planetesimals being spread across a wider size range, hence more planetesimals will be generated. Here we choose $q=-3$, instead of the typical value $q=-3.5$, as it will increase the number of planetesimals generated. Using a density of $3\, \rm{g\, cm^{-3}}$, the non-grouped planetesimal mass is used to generate a distribution of non-grouped planetesimals according to the set size distribution, and an upper size set by the smallest grouped planetesimal. Overall, we have three populations of bodies: 1) largest remnant (plus second largest remnant if the criterion is met), 2) grouped planetesimals defined from groupings of escaping particles in the SPH output that fill out the large mass end of the planetesimal size distribution, and 3) the non-grouped planetesimals which have an assumed power law size distribution with the total mass derived from the SPH output that fill out the bottom end of the size distribution. 

\begin{figure*}
    \centering
    \includegraphics[width=\linewidth]{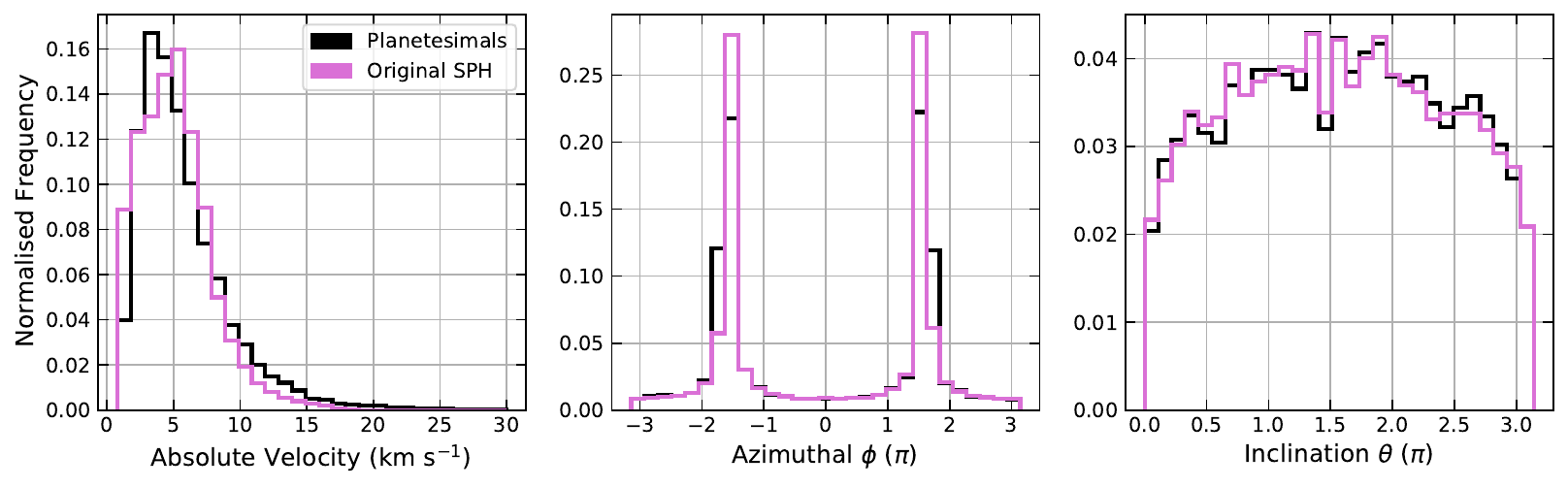}
    \caption{The distribution of absolute velocities (left), azimuthal $\phi$ (middle), and inclination (right) of the distribution of particles in the giant impact frame of reference. The escaping particles in the SPH giant impact simulation are shown in purple and the planetesimals formed from the escaping particles are shown in black. The escaping particles and planetesimals shown are from giant impact 1 in Table \ref{tab:giant_impacts}.}
    \label{fig:plan_sph_compare}
\end{figure*}

Finally, velocity kicks are given to the planetesimals from the impact site.
For the grouped planetesimals, the velocity kick is calculated from the SPH simulation by tracking the centre of mass velocity magnitude and direction of the bound group of SPH particles they were defined from. The non-grouped planetesimals will follow the distribution of the unbound escaping particles from the SPH simulations. A mean and standard deviation are calculated for the unbound escaping particles in log space which defines a log-normal distribution from which the absolute velocity of the unresolved planetesimals is drawn. The direction of the kick is determined through randomly sampling the distribution of angles $\phi$ and $\theta$ of the unbound escaping particles from the impact, where $\phi$ is the angle measured in the x--y plane of the SPH simulation from the COM of the largest remnant (or two largest remnants), and $\theta$ is the angle measured in the xy--z plane. Fig. \ref{fig:giant_impact_example} shows an impact in the x--y plane. 
Fig. \ref{fig:plan_sph_compare} shows the distribution of $v, \phi\ \rm{and}\ \theta$ for the full planetesimal distribution (grouped and non-grouped) in black and the escaping SPH particles of giant impact 1 in purple. We find the planetesimal distribution of $\phi$ and $\theta$ to match well to the escaping SPH particles but we note there is a slight difference in the velocity distribution. The difference is down to the unbound escaping SPH particles not having a symmetric log-normal distribution. Hence, while the mean of both distributions are the same, the median values are slightly different, with our generated planetesimals having a slightly smaller median value. For our study, the differences between the velocity distributions will have a minuscule effect on the overall result as the direction of the kick is what drives a large difference. 

\begin{table*}
\caption{Summary of giant impact set up and results from SPH simulations. $\rm{M_{tot}}-$ total mass in impact in Earth masses; $\rm{M_{proj}/M_{targ}}-$ mass ratio of projectile to target mass; $\rm{N}-$ number of particles used in the simulation; $v_i-$ impact velocity in $\rm{km\ s^{-1}}$; $b-$ the impact parameter; $\rm{M_{lr}/M_{tot}}-$ mass ratio of the largest remnant to total mass in the simulation; $\rm{M_{sr}/M_{tot}}-$ mass ratio of the second largest remnant to the total mass in the simulation; $M_{\rm{plan }}-$ total planetesimal mass formed in Earth masses.}
\label{tab:giant_impacts}
\begin{tabular}{ c c c c c c c c c }
\hline
Index & $\rm{M_{tot}}$ & $\rm{M_{proj}/M_{targ}}$ & N & $v_i$ & $b$ & $\rm{M_{lr}/M_{tot}}$ & $\rm{M_{sr}/M_{tot}}$ & $M_{\rm{plan}}$ \\
$-$ & ($\rm{M_{\oplus}}$) & $-$ & ($10^4$) & ($\rm{km\ s^{-1}}$) & $-$ & $-$ & $-$ & ($\rm{M_{\oplus}}$)  \\
\hline\hline
1 & 2.37e-01 & 1.00 & 40 & 10.0 & 0.00 & 0.697 & 0.007 & 5.82e-02\\
\hline
2 & 3.71e-01 & 0.47 & 40 & 11.4 & 0.42 & 0.654 & 0.195 & 4.13e-02\\
\hline
3 & 1.99e-01 & 1.00 & 4 & 15.0 & 0.00 & 0.376 & 0.000 & 5.83e-02\\
\hline
\hline \\ 
\end{tabular}
\end{table*}

\subsection{N-body}

To evolve the post-impact planetesimals, we used {\sc rebound} \citep{Rein_&_Liu_2012_main_rebound}, an all-purpose $N$-body code which allows for collision detection. We chose the collision detection Linetree method which checks to see if two particles overlapped between two timesteps. Linetree works best for particles paths between steps being straight lines therefore we chose a leapfrog integrator to evolve our simulations. In our $N-$body simulations we vary the initial giant impact position between 0.3 au and 2 au around a star of 1 \som, with the progenitor orbit (the orbit of the target planet in the giant impact) set to always be circular. The time-step we used between steps was 1/2000 of the orbital period of the progenitor orbit as early on all the planetesimals are clumped around the embryo remnant(s). A larger timestep would lead to an inaccuracy with calculating the trajectories of the particles near the beginning of the simulation. We also found an insufficiently small timestep would lead to the collision point widening on shorter timescales leading to fewer collisions occurring. 

The planetesimals are evolved for $10^4$ orbits of the progenitor orbit with each simulation having a particle number resolution defined by the resolved planetesimals and unresolved planetesimal mass from the SPH simulations. The complete set of $N-$body simulations can be found in Table \ref{tab:disk_table}. 
We assume that the planetesimals are non-gravitating, meaning the planetesimals only interact gravitationally with the central star and the planetary embryo remnant(s). For the larger planetesimals, gravitational focusing may play a role in expanding their collision cross-section therefore our assumption might underestimate the overall collision rate.
A lack of gravitational interaction may lead to large errors at early times before the disk has fully stirred through missed collisions. Though in all simulated disks we find the smallest mean relative velocity to be $\sim 2 \rm{\ km\ s^{-1}}$ which is greater than the escape velocity of the largest planetesimal in all simulations $\sim 1.2 \rm{\ km\ s^{-1}}$. The mean relative velocity between planetesimals increases quickly after the start of the simulations. Hence, we can ignore gravitational focusing as a large source of error.
Once a collision is detected, we determine the catastrophic impact velocity $V^*$, the velocity at which half the mass is disrupted, using methods outlined in \citet{Leinhardt-2012-collisions-between-gravity-dom-bodies},

\begin{equation*}
\centering
    V^* = \left[\frac{1}{4}\frac{(\gamma+1)^2}{\gamma}\right]^{1/(3\bar{\mu})}V^*_{\gamma=1},
\end{equation*}
where,
\begin{equation*}
\centering
    V^*_{\gamma=1} = \left(\frac{32\pi c^*\rho_1 G}{5}\right)^{1/2}R_{C1}.
\end{equation*}
Here, $\gamma$ is the mass ratio of the two colliding planetesimals, $R_{C1}$ is the combined radius given a density $\rho_1 = 1000\ \rm{kg\ m^{-3}}$, $\bar{\mu}$ and $c^*$ are fit parameters determined to be 0.37 and 5 respectively for small bodies in \citet{Leinhardt-2012-collisions-between-gravity-dom-bodies}.
After calculating $V^*$ we determine collision outcomes. We assumed a simplified collision outcome criteria in order to save on computational time. The collision outcome criteria is based on the impact speed, $v_i$, and has three outcomes: 1) $v_i/V^*<0.1$ the planetesimals merge, 2) $0.1 \leq v_i/V^* < 2$ results in a bouncing collision, and 3) $v_i/V^*\geq2$ the collision is completely destructive and both planetesimals are destroyed. We allow the clump to initially evolve for 0.25 orbits before allowing collisions.
Our study focuses on the visible debris formed from planetesimal collisions post-giant impact. Tracking the most destructive collisions means that we are tracking the quickest path that the planetesimal mass has on influencing the debris disk formed after a giant impact. However, we will miss mass from partially erosive collisions over time so only tracking the most destructive collisions will lead to an underestimate of the mass which will grow over time. The collision outcome does not take into account impact angle, hence all collisions are assumed to be head-on. The result of this assumption means greater numbers of destructive impacts than would otherwise occur. As steps are taken in straight lines, along with collision detection, the impact angle will differ from a more complicated integrator. Since the impact angle will vary from its true value, we decided to take the most extreme assumption with head-on collisions. The reasoning being that if an observable disk is not formed through this favourable set-up, then impact angle will not matter. The planetesimals are usually large with the smallest non-grouped planetesimal generated from giant impact 1 having a radius of 49.6 km, meaning they are well within the gravity regime and not the strength regime \citep{1994P&SS...42.1067H,1996LPI....27....1A}. Therefore, using the collisional outcome criteria we have set and adopting the catastrophic impact velocity from \citet{Leinhardt-2012-collisions-between-gravity-dom-bodies} is justified. 

\section{Results}
\label{sec:results}

An extreme debris disk formed from the escaping debris of a giant impact will vary depending on many factors. The initial dust formed from vaporised material is only expected to survive for a few orbits before being ejected through radiation pressure or other means \citep{Su-2019-extreme-disk-variability}. 
Though we find that extreme debris disks can be observable for many orbits after a supposed giant impact has occurred. We showed in \citet{Watt_2021_vapour_EDD} that the vapour condensate formed initially after a giant impact can reproduce behavior and the flux increase in EDDs. \citet{Su-2019-extreme-disk-variability} suggested that ID8 in 2017 had a steady increase in the excess flux associated with a collisionally active boulder/planetesimal population. In this work we aim to understand the collisional evolution of planetesimals formed after the giant impact to explain the extended lifetimes of EDDs. 
Here we focus on how a planetesimal population formed from the escaping ejecta of a giant impact can sustain an EDD through a collisional cascade. By generating a planetesimal population from SPH simulations of giant impacts we can numerically study the collisional evolution of the post-giant impact planetesimal disk through $N$-body simulations.

\subsection{N-body}

\begin{figure*}
    \centering
    \includegraphics[width=\linewidth]{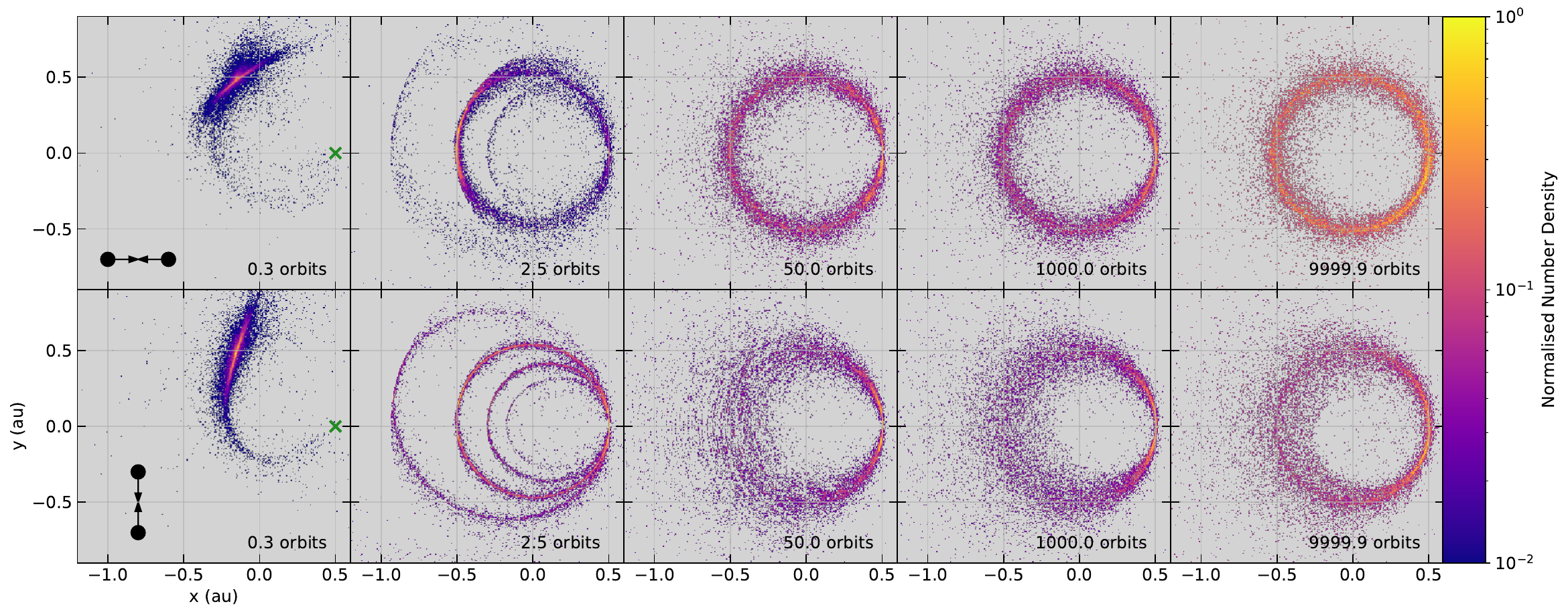}
    \caption{Evolution of disks 3 (top) and 13 (bottom) from Table \ref{tab:disk_table} with 5 snapshots after 0.3, 2.5, 50, 1000, and 9999.9 orbits of the progenitor orbit post-giant impact. Both disks formed from giant impact 1 from Table \ref{tab:giant_impacts}. The green cross marks the location of the giant impact in the first panel. The diagram in the bottom left shows the orientation of the giant impact. The colour denotes the normalised number density in each panel, i.e. the maximum number density value in each panel differs.}
    \label{fig:nbody_snapshots}
\end{figure*}

We simulated a total of 64 collisionally active disks for 3 different giant impacts. Most disks simulated focus on giant impact 1, with the other two impacts offering an insight into another parameter space. Fig. \ref{fig:nbody_snapshots} shows five snapshots of two planetesimal disks placed at 0.5 au formed from giant impact 1. The evolution of the disk is similar to that seen in \citet{Jackson2012_Moon,Jackson-2014-planetary-collisions-at-large-au,Watt_2021_vapour_EDD}. It is expected that the distribution shown in fig \ref{fig:nbody_snapshots} is similar to what is seen in \citet{Watt_2021_vapour_EDD} as most planetesimals have velocity kicks defined from the escaping ejecta the same as for the escaping vapour condensates. 


\subsubsection{Impact Velocity Distribution}
\begin{figure}
    \centering
    \includegraphics[width=\linewidth]{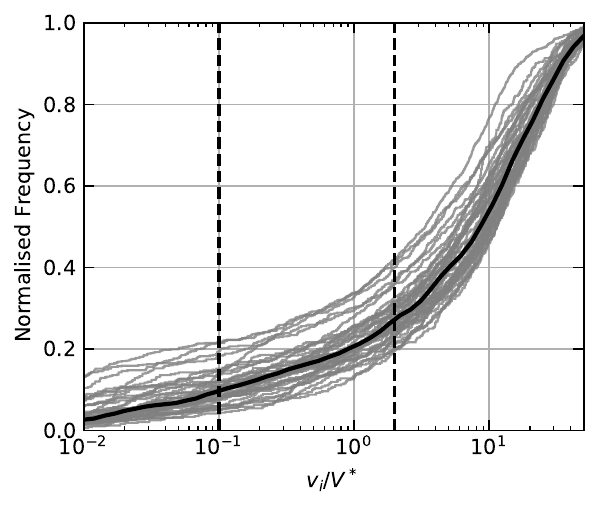}
    \caption{The cumulative distribution of $v_i/V^*$ for disks 13 to 52 in Table \ref{tab:disk_table}. The black line shows the median interpolated distribution. The black dashed lines indicate the conditions for the collision outcomes.}
    \label{fig:norm_freq_vel_ratio_no_rem_many}
\end{figure}

In our simplified collisional outcome prescription, the mass which is destroyed in planetesimal-planetesimal collisions is the mass that will form into a range of grain sizes that are observable. The complete destruction of planetesimals requires $v_i>2V^{*}$. In Fig. \ref{fig:norm_freq_vel_ratio_no_rem_many} we show the normalised frequency of the impact velocities in planetesimal-planetesimal collisions normalised by the catastrophic impact velocity for 40 simulated disks formed from giant impact 1 with the impact having an orientation of $0.5\pi$ at 0.5 au. How the orientation of a giant impact can vary the disk structure is shown in fig. \ref{fig:nbody_snapshots}, the orientation refers to the direction of impact with respect to the progenitor orbit. An impact with $0\pi$ orientation has the impact occur in the same direction as the progenitor orbital velocity vector. The black line in fig. \ref{fig:norm_freq_vel_ratio_no_rem_many} shows the median distribution. For the median distribution, we find that 73 per cent collisions are destructive with the value ranging from 63 per cent to 84 per cent in individual runs. The range in the number of destructive collisions is caused by the random sampling of velocities when creating the distribution of planetesimals from the escaping ejecta in the giant impact. The difference seen between the percentage of destructive impacts can be explained by $V^*$ having a mass and mass ratio dependence. More massive planetesimals will require either a faster impact speed and/or to impact into a more similarly sized planetesimal. 

\begin{figure}
    \centering
    \includegraphics[width=\linewidth]{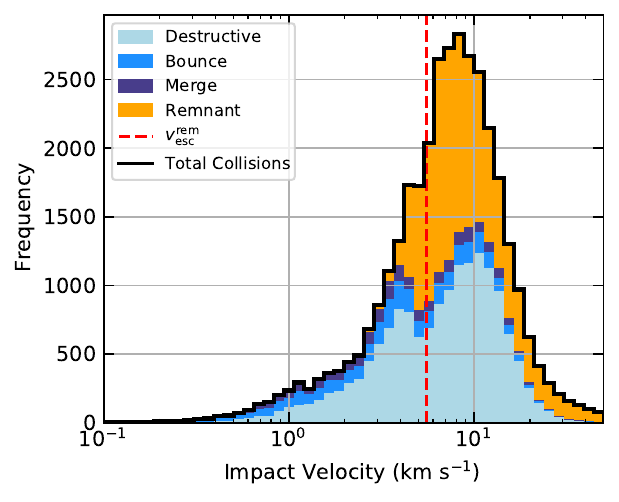}
    \caption{The distribution of impact velocities of planetesimal-planetesimal collisions and remnant-planetesimal collisions. The histogram counts all collisions for disks 13 to 52 in Table \ref{tab:disk_table}. The colour denotes the frequency of a type of collision, blues are for planetesimal-planetesimal collisions and orange for remnant-planetesimal collisions. The blues split into: light blue for destructive outcome, blue for bounce outcome, and purple for merge collisions. All remnant-planetesimal collisions are merging. The red dashed line represents the escape velocity of the remnant, $v_{\rm{esc}}^{\rm{rem}}$.The black line represents the total number of collisions.}
    \label{fig:imp_vel_hist_all_stacked}
\end{figure}

The mass of the planetesimals is set by the size distribution, therefore there are far more lower mass planetesimals of similar sizes than larger planetesimals. With smaller planetesimals being more numerous and requiring less energy in the impact to be disrupted, there is a greater chance that smaller planetesimals are destroyed in the simulations. Fig. \ref{fig:imp_vel_hist_all_stacked} shows the stacked distribution of all collisions recorded in each of the 40 simulations for giant impact 1 with orientation of $0.5\pi$ at 0.5 au. The stacked histogram splits the collision type for each planetesimal-planetesimal impact, and remnant-planetesimal impacts are always merging due the remnant being much larger. We find that most impacts between planetesimals are relatively fast, above $3\ \rm{km\ s^{-1}}$, which can destroy two planetesimals with the same size of $\sim 680$ km. We do not see a uniform distribution however, the distribution looks to be more bimodal. The peak between $3-4\ \rm{km\ s^{-1}}$ is caused by collisions within the first orbit after the giant impact, while 85.4 per cent collisions occur at the collision point after one orbit. The dip in planetesimal-planetesimal collisions occurs at the expected escape velocity of the remnant. The link between the escape velocity and the bimodal distribution can be down to the velocity kicks given to the planetesimals not being sufficient to clear the area around the remnant before reaching 0.25 orbits when collisions start to be resolved in the simulations.

\subsubsection{Collisions and Mass}

\begin{figure}
    \centering
    \includegraphics[width=\linewidth]{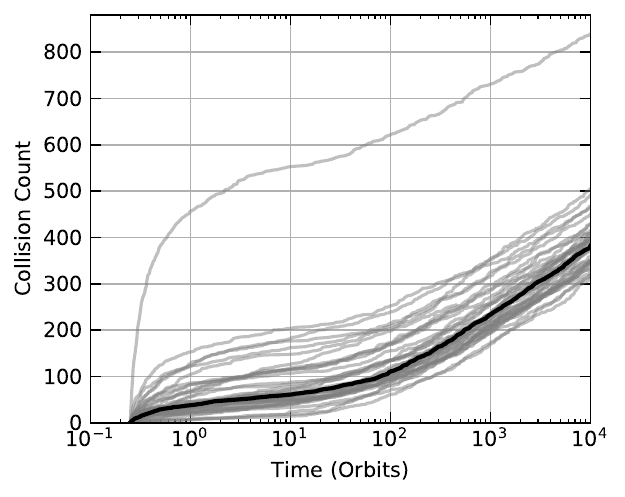}
    \caption{The destructive collision count over $10^4$ orbits for disks 13 to 52 in table \ref{tab:disk_table}. The grey lines represent individual simulations and the black line represents the median interpolated collision count.}
    \label{fig:des_col_count_all}
\end{figure}

\begin{figure}
    \centering
    \includegraphics[width=\linewidth]{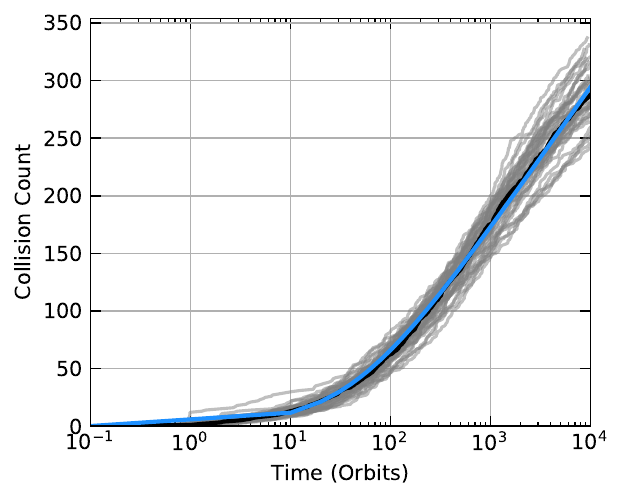}
    \caption{Same as Fig. \ref{fig:des_col_count_all} but only counts destructive collisions within $|\phi| < \pi/10$ azimuthal slice of the collision point. The blue line is the fit to the data using equation (\ref{eqn:colp_count}) with $a=53.54$ and $t_p = 41.22$. }
    \label{fig:des_col_count_colp}
\end{figure}

The number of collisions in a simulation will indicate how active each disk is. In Fig. \ref{fig:des_col_count_all} we show the cumulative number of destructive collisions over time in orbits for disks 13 to 52 in Table \ref{tab:disk_table} in grey and the median cumulative count in black. We see an initial flurry of destructive collisions early on before moving into a steadier increase in destructive collisions over time. We find that there are two different causes of collisions between planetesimals: the early time collisions, which mostly take place soon after collisions are switched on in the simulations (0.25 orbits), and collisions that take place around the collision point. The collision point is defined as the location where the initial giant impact occurred. For collisions taking place at the collision point, we find the total number of collisions varies over time as
\begin{equation}
\label{eqn:colp_count}
        N_{\rm{tot}}(t) =  a{\rm{log}}\left(1 + \frac{t}{t_p}\right),
\end{equation}
where $t_p$ is the turning point, and $a$ being a constant. Fig. \ref{fig:des_col_count_colp} like Fig. \ref{fig:des_col_count_all} shows the destructive collision count but only for an azimuthal slice of $|\phi| < \pi/10$ around the collision point. The increase in destructive collision count at the collision point over time is more uniform than when accounting for all the impacts. The less uniform distribution in Fig \ref{fig:des_col_count_all} suggests that the randomness of the planetesimal setup can cause large differences in collision outcomes within the first few orbits away from the collision point. 

\begin{figure}
    \centering
    \includegraphics[width=\linewidth]{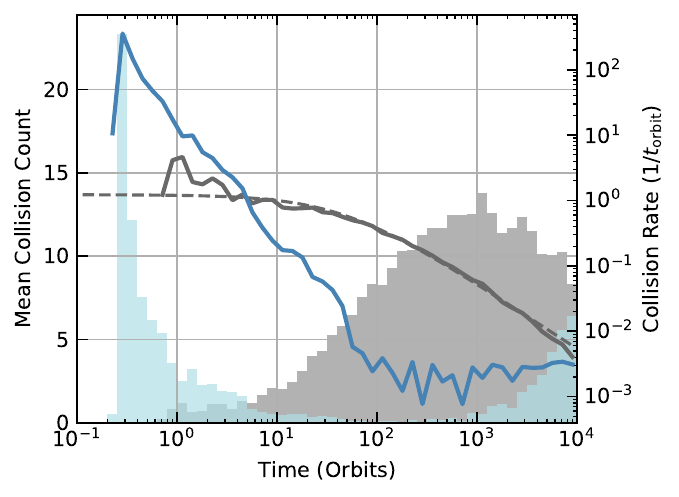}
    \caption{Left y-axis: histogram of the mean count of collisions in disks from 13 to 52 in Table \ref{tab:disk_table}. Right y-axis: the mean collision rate with the solid lines determined from the mean count in each histogram bin divided by the bin width. The data is split between collisions occurring in an azimuthal slice of $|\phi| < \pi/10$ around the collision point (grey) and collisions occurring elsewhere (blue). The black dashed line represents the expected collision rate for collisions around the collision point from equation (\ref{eqn:des_colp_col_rate}) with parameters fit from the median collision frequency in Fig.\ref{fig:des_col_count_colp}.}
    \label{fig:des_col_count_compare}
\end{figure}

\begin{figure}
    \centering
    \includegraphics[width=\linewidth]{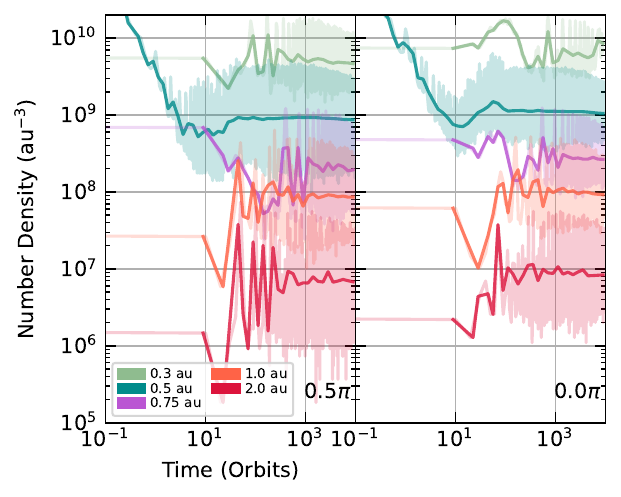}
    \caption{The number density of particles within 10 Hill radii of the remnant over time. The solid lines show binned density while the faint solid lines show the full data. The colour of lines represents the semi-major axis of the giant impact location (light green: 0.3 au, dark green: 0.5 au, purple: 0.75 au, orange: 1 au, red: 2 au). The disks were formed from giant impact 1 with an orientation of $0.5\pi$ (left) and $0.0\pi$ (right). The disks all had the same initial velocity distribution and only differed in location and giant impact orientation. Disks represented in plot can be found in Table \ref{tab:disk_table} and are listed starting from the top downwards as follows: 2, 13, 54, 56, and 58 on the left, and 1, 3, 53, 55, and 57 on the right. Disks 3 and 13 placed at 0.5 au is the only disks to be shown to extend below 10 orbits as they are the only disks for which output steps were saved every 0.1 orbits, the rest had steps saved every 10 orbits.}
    \label{fig:rem_density}
\end{figure}

The collision count as a function of time relation in equation (\ref{eqn:colp_count}) means we should expect the collision rate to vary over time as 
\begin{equation}
    \label{eqn:des_colp_col_rate}
    R_{\rm{col}}(t) = \frac{a}{t_p + t}.
\end{equation}
Note we only consider the most extreme collisions to be destructive collisions, hence $R_{\rm{col}(t)}$ varies over time mostly through dynamical evolution not collisional. There will be mass loss from planetesimal collisions that fall below our criteria for destructive collisions that is not accounted for. The mean collision rate is shown in Fig. \ref{fig:des_col_count_compare}. The grey bins are the destructive collision counts around the collision point and the grey line is the collision rate calculated through dividing the collision count in each bin by the bin width. The blue bins and line represent the collision count and collision rate for collisions away from the collision point. The black dashed line shows the expected collision rate from equation (\ref{eqn:des_colp_col_rate}) with parameters fit from the median collision frequency shown in Fig.\ref{fig:des_col_count_colp}. We show that the expected collision rate is a good fit for collisions occurring at the collision point from five orbits onwards. The deviations away from the fit before five orbits could be caused by the remnant disturbing a large selection of planetesimals. As the planetesimals are escaping from the remnant at the start of the simulations, the remnant will have a large influence over the dynamics of the planetesimals in the first few orbits while the disk is starting to form. We see in Fig. \ref{fig:rem_density} that the density inside 10 Hill radii of the remnant decreases sharply over the first five orbits of the simulations for disk 13 (teal) before remaining relatively constant throughout the rest of the simulation. The planetesimal number density decreasing rapidly around the remnant could be related to the collision rate dropping rapidly below the collision rate around the collision point. However, the collision rate outside the collision point keeps decreasing until around 100 orbits after the giant impact, much later than the density around the remnant levels off. 

\begin{figure}
    \centering
    \includegraphics[width=\linewidth]{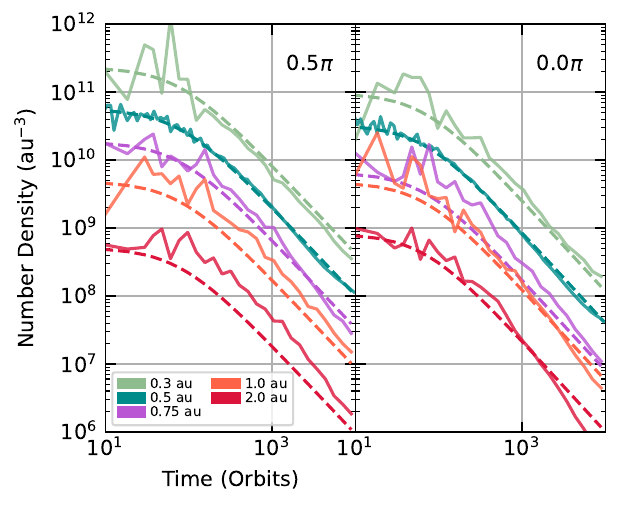}
    \caption{The number density at the collision point over time. The solid lines are binned mean number density at the collision point. The dashed lines are the expected number density evolution fitted using equation (\ref{eqn:number_density}) with fit parameters [left, right] of $t_{\rm colp} = [77,\ 66]$ and $p$ = [1.3, 1.34]. The colour of lines represents the semi-major axis of the giant impact location (light green: 0.3 au, dark green: 0.5 au, purple: 0.75 au, orange: 1 au, red: 2 au). The disks were formed from giant impact 1 with an orientation of $0.5\pi$ (left) and $0.0\pi$ (right). The disks all had the same initial velocity distribution and only differed in location and giant impact orientation. Disks represented in plot can be found in Table \ref{tab:disk_table} and are listed starting from the top downwards as follows: 2, 13, 54, 56, and 58 on the left, and 1, 3, 53, 55, and 57 on the right. The parameter $b$ is set through calculating the median value when $t << t_{\rm colp}$, hence there are only a limited number of data points to calculate $b$ besides for disks 3 and 13.}
    \label{fig:number_density_sma}
\end{figure}

For collisions that occur around the collision point, we find correlation between the decrease in the number density of planetesimals at the collision point and the number of collisions recorded overall. 
To measure the number density we randomly sample the space around the collision point. We choose to sample up to $0.05 a_{\rm sma}$ away from the collision point, where $a_{\rm sma}$ is the semi-major axis of the disk. The choice of a dynamic distance instead of using a fixed distance away from the collision point is to take into account how the disk structure scales with sma. We randomly sample the space $10^5$ times and calculate the volume of a sphere for each random sample, $V_{\rm sample}$, where the radius is the distance to the $5^{\rm th}$ closest $N$-body particle. The number density of the collision point is then calculated as $n_{\rm colp} = 5/\rm{min}(V_{\rm sample})$.
In Fig. \ref{fig:number_density_sma} we show the number density around the collision point for simulated disks at varying semi-major axes and two different giant impact orientations. We discuss how the position of the disk and orientation affects disk evolution in sections \ref{sec:SMA} and \ref{sec:orientation}. For now we focus on disk 13 (teal, left panel) which has an semi-major axis of 0.5 au and is formed from a giant impact with an orientation of $0.5\pi$. The number density is found to vary over time as,
\begin{equation}
    \label{eqn:number_density}
    n_{\rm{colp}}(t) = \frac{b}{\left(1 + \frac{t}{t_{\rm colp}}\right)^p},
\end{equation}
where $t_{\rm colp}$ is the turning point, with $b$ and $p$ being constants. We find that disk 13 can be fitted approximately with the same turning point value for both equations (\ref{eqn:colp_count}) and (\ref{eqn:number_density}). For the density fit, we used a value of $t_{\rm colp} = 77$ orbits. The decrease in number density of planetesimals at the collision point decreases the collision rate over time after an initial steady period. The collision point starts to smooth out over a timescale of over 100 orbits for a planetesimal disk formed from giant impact 1 with an orientation of $0.5\pi$ at 0.5 au. 

\begin{figure}
    \centering
    \includegraphics[width=\linewidth]{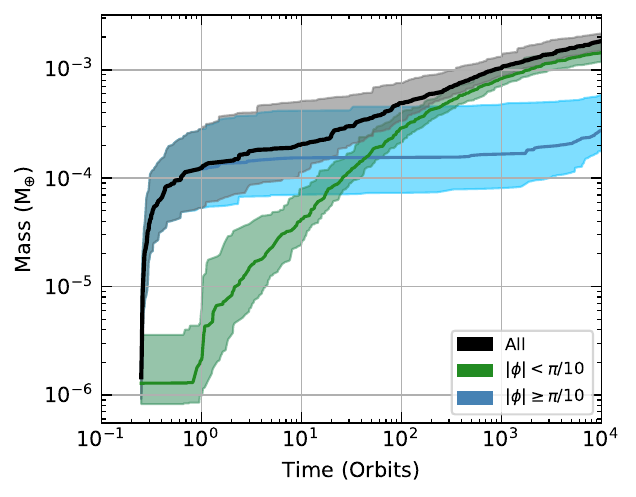}
    \caption{The median mass produced in destructive collisions in disks 13 to 52 in Table \ref{tab:disk_table}. All collisions are shown in black, while collisions that occur within an azimuthal cut of $|\phi| < \pi/10$ around the collision point and collisions outside the collision point are shown in green and blue respectively. The coloured regions represent the 16th and 84th percentiles of the mass.}
    \label{fig:des_mass_all}
\end{figure}

We track the collisions to get an estimate for the mass that will be passed down the size distribution to grains which will impact the observability of the extreme debris disk in near to mid infrared wavelengths. Fig. \ref{fig:des_mass_all} shows the median disk mass from disks 13 to 52 in Table \ref{tab:disk_table} over $10^4$ orbits in black, and also shows the median mass produced within the azimuthal cut of $|\phi| < \pi/10$  and the mass produced outside the azimuthal cut in green and blue respectively. The coloured area around each line shows the 16th and 84th percentiles of mass produced.  We see that the collisions between planetesimals outside the collision point give a sharp increase in mass in the initial few orbits. The mass produced at the collision point does come to dominate after $\sim$100 orbits and by $10^4$ orbits the mass produced is an order of magnitude more than that produced outside the collision point, though we see in Fig. \ref{eqn:des_colp_col_rate} that the collision rate around the collision point decreases over time. As we expect the disk to become more symmetric over time, the collision rate around the collision point will match the collision rate around the rest of the disk. Hence, at late times past the $10^4$ orbits we simulate we would expect the contribution from the collision point to the mass produced to be less than that from the rest of the disk. The collision point enhances the number of collisions that we would expect from a symmetric disk in the disk's early formation and evolution. 

\subsubsection{Semi-major axis variation}
\label{sec:SMA}

\begin{figure}
    \centering
    \includegraphics[width=\linewidth]{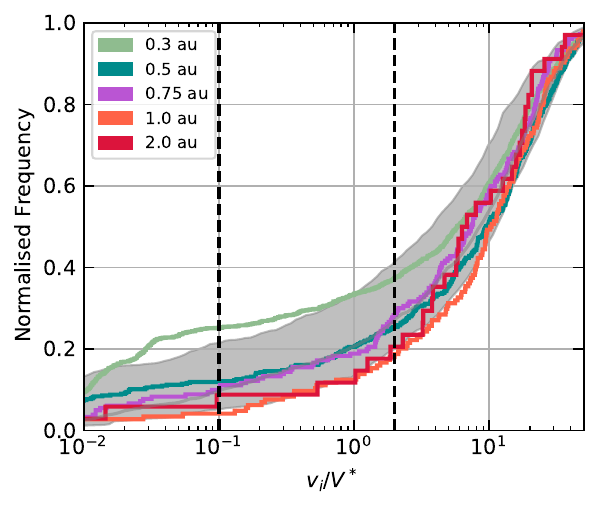}
    \caption{The cumulative velocity distribution for disks formed from giant impact 1 with an orientation of $0.5\pi$ at different semi-major axes. The disks are disks 2 (light green, 0.3 au), 13 (dark green, 0.5 au), 54 (purple, 0.75 au), 56 (orange, 1 au), and 58 (red, 2 au) in Table \ref{tab:disk_table}. The grey denotes the 2.5th and 97.5th percentile of the median velocity distribution shown in Fig. \ref{fig:norm_freq_vel_ratio_no_rem_many}. }
    \label{fig:vel_distrib_au}
\end{figure}

Giant impacts do not only occur at 0.5 au, but can occur at any distance from their host star. For giant impact 1 with orientation of $0.5\pi$, we vary the semi-major axis (sma) at which the impact occurred. We choose sma values of 0.3, 0.5, 0.75, 1 and 2 au with the disks being listed in Table \ref{tab:disk_table} as disks 2, 13, 54, 56, and 58. We limit the choice of sma values to 2 au as while giant impacts can occur further out, extreme debris disks are observed within the terrestrial planet formation zone. The disks use the same random seed for obtaining the properties of the planetesimals. The $v_i$/$V^*$ distribution is shown in Fig. \ref{fig:vel_distrib_au} for the giant impact induced disks at different sma. The coloured lines represent the different sma, and the grey shows the spread between the 2.5th and 97.5th percentile of the velocity distribution shown in Fig. \ref{fig:norm_freq_vel_ratio_no_rem_many}. We find that while there appears to be some difference in the distribution going towards more destructive collisions at larger sma, all lines fall within the random sampling of a disk placed at 0.5 au. There is no major difference found between these single runs in terms of collisional outcomes. The collisional outcomes are expected to be similar as the disks are formed from the same giant impact. While we vary the bulk Keplerian motion given to the planetesimals, the velocity dispersion of the planetesimal population will remain the same. If the velocity dispersion remains the same then there should be no significant difference in relative speeds of the planetesimals. It is possible that there could be a difference when more runs are conducted but there would still have to be some overlap. The small difference seen between the distributions could be down to the number of collisions occurring in each disk. 

\begin{figure}
    \centering
    \includegraphics[width=\linewidth]{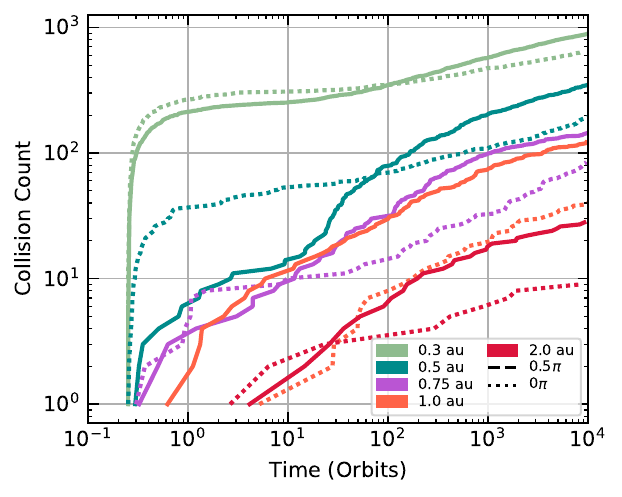}
    \caption{The cumulative collision count for disks formed from giant impact 1 with an orientation of $0.5\pi$ (solid) and $0\pi$ (dashed). The disks shown are disks 2 (light green, 0.3 au), 13 (dark green, 0.5 au), 54 (purple, 0.75 au), 56 (orange, 1 au), and 58 (red, 2 au) in table \ref{tab:disk_table} over $10^4$ orbits.}
    \label{fig:col_freq_au}
\end{figure}

The number of collisions for the disks at different sma are shown in Fig. \ref{fig:col_freq_au}. The number of collisions decreases as we move away from the star. We also see the initial collisional activity disappear beyond 0.5 au. The initial collisional activity dropping off can be explained by the change in the bulk Keplerian velocity given to the planetesimals. The Keplerian velocity varies as $v_k \propto r^{-\frac{1}{2}}$ for a circular orbit, where $r$ is the distance. The ratio of velocity dispersion to Keplerian velocity then varies as $\sigma_v/v_k \propto r^{\frac{1}{2}}$, if $\sigma_v$ is kept constant. For the same giant impact we would expect the disk to shear out on a faster orbital timescale further away from the star. Hence when collisions are turned on at 0.25 orbits, the planetesimals around the remnant have sheared out quicker in disks placed at greater sma. We see in Fig. \ref{fig:rem_density} how the number density close to the remnant decreases as you move further away from the star, hence the remnant dynamically influences fewer planetesimals. 

When the collision count is normalised, equation (\ref{eqn:colp_count}) still holds for collisions occurring at the collision point. This suggests that the orbital timescale for the collision point to smooth out is the same regardless of where the giant impact occurred. Though we do see a difference in the number of collisions that happen outside the collision point due to $\sigma_v/v_k$ scaling with distance from the star. The close in disks have a planetesimal population that is more densely packed around the remnant in the initial few orbits. 

The difference in the number of collisions between planetesimals can be explained by the difference in density at the collision point. The collision point does evolve the same for all disks but the initial density differs. We see in Fig. \ref{fig:number_density_sma} how the density at the collision point changes over time for disks from 0.3 to 2 au. The density of particles at the collision point directly relates to the number of collisions expected. Since all disks share the same number of particles, it is likely that the initial volume of the collision point is larger for disks further out reducing the density. Again, the change in $\sigma_v/v_k$ is the likely cause.

\begin{figure}
    \centering
    \includegraphics[width=\linewidth]{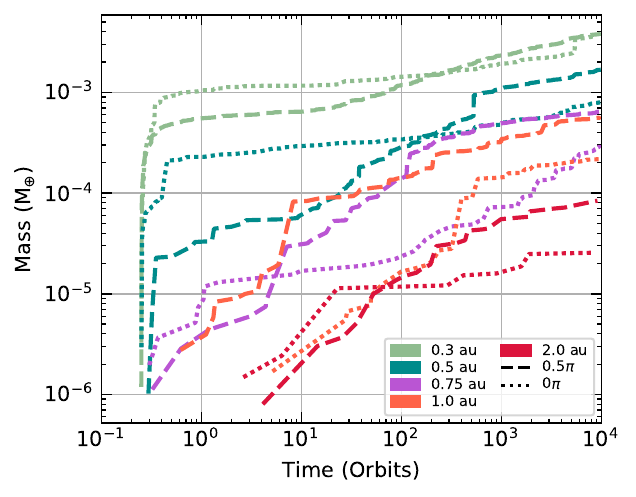}
    \caption{Mass produced through destructive planetesimal collisions over time for the disks formed from giant impact 1 with orientations $0\pi$ (dotted) and $0.5\pi$ (dashed) between 0.3 au and 2 au. The disks shown are disks 2 (light green, 0.3 au), 13 (dark green, 0.5 au), 54 (purple, 0.75 au), 56 (orange, 1 au), and 58 (red, 2 au) in table \ref{tab:disk_table} over $10^4$ orbits.}
    \label{fig:disk_mass_au}
\end{figure}

\begin{figure}
    \centering
    \includegraphics[width=\linewidth]{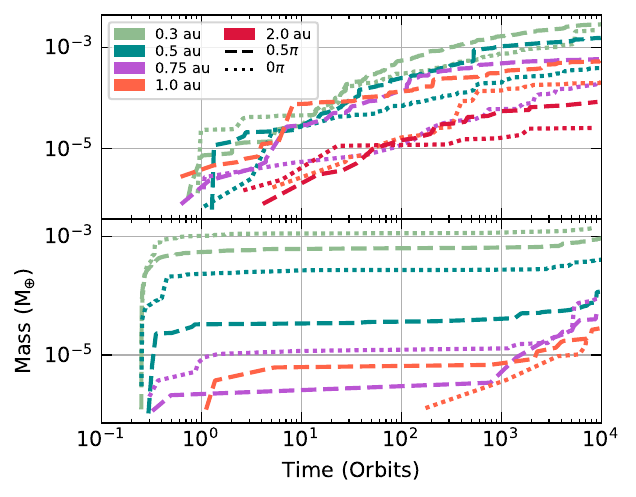}
    \caption{Disk mass split into mass produced within a azimuthal cut of $|\phi|<\pi/10$ around the collision point (top) and outside the collision point (bottom). Disks are formed from 0.3 au to 2 au. The disks shown are disks 2 (light green, 0.3 au), 13 (dark green, 0.5 au), 54 (purple, 0.75 au), 56 (orange, 1 au), and 58 (red, 2 au) in table \ref{tab:disk_table} over $10^4$ orbits.. Dashed lines are for disks formed from giant impact 1 with an orientation of $0.5\pi$ and dotted lines are for an orientation of $0\pi$. }
    \label{fig:disk_mass_au_split}
\end{figure}

The mass of debris produced through planetesimal collisions for each disk is shown in Fig. \ref{fig:disk_mass_au}. The sma are varied by colour, the dashed lines show the mass for disks formed from giant impact 1 with orientation $0.5 \pi$, and the dotted lines show the mass for the same impact with an orientation of $0 \pi$. We find between the 0.3 au and 2 au case over an order of magnitude difference in the mass of the disk. The other disks fall within this range. We see in Fig. \ref{fig:disk_mass_au_split} the mass over time of each disk split into mass produced around the collision point (top) and mass produced from collisions elsewhere (bottom). The disks close into the star experience a large contribution to the debris mass from collisions outside the collision point within the first few orbits. Beyond the 0.5 au disk we do not see a significant contribution to mass. Of course these are only one potential snapshot of the random distributions the planetesimals can take on. We might see a significant contribution from disks outside 0.5 au in some snapshots though this seems unlikely when comparing the densities around the remnant between the disks. 

\begin{figure}
    \centering
    \includegraphics[width=\linewidth]{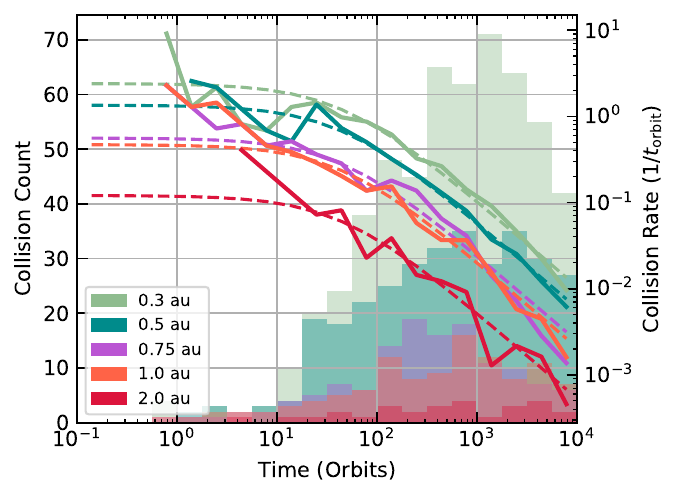}
    \caption{Left: An histogram of collisions in disks 2 (light green, 0.3 au), 13 (dark green, 0.5 au), 54 (purple, 0.75 au), 56 (orange, 1 au), and 58 (red, 2 au) in table \ref{tab:disk_table} over $10^4$ orbits. Right: Collision rate for each disk calculated from the bin count divided by the bin width. Dashed lines show fits for the collision rate using equation (\ref{eqn:des_colp_col_rate}).}
    \label{fig:des_col_rate_au}
\end{figure}

The effect of the lower density is seen in Fig. \ref{fig:des_col_rate_au}. The behaviour of each disk is the same over time, the collision rate decreases following equation (\ref{eqn:des_colp_col_rate}). It is the initial collision rate that varies with distance from the star. The further away the giant impact occurs, the smaller the initial collision rate. 

\subsubsection{Orientation}
\label{sec:orientation}

All giant impacts distribute ejecta anisotropically. The orientation which the giant impact occurs at with respect to the stellar system reference frame can have a huge impact on the disk structure. We showed in \citet{Watt_2021_vapour_EDD} how the orientation affects the light curve seen from a vapour condensate disk, specifically the dips in the light curve at the collision point and anti-collision line not always appearing. For giant impact 1 we ran 40 simulations for disks forming at 0.5 au formed with an impact orientation of $0.5\pi$, we have also run 10 simulations for an orientation of $0\pi$ to understand if orientation plays a large role in planetesimal collisions as it does in the behaviour of the light curve in the vapour condensate disk. 

\begin{figure}
    \centering
    \includegraphics[width=\linewidth]{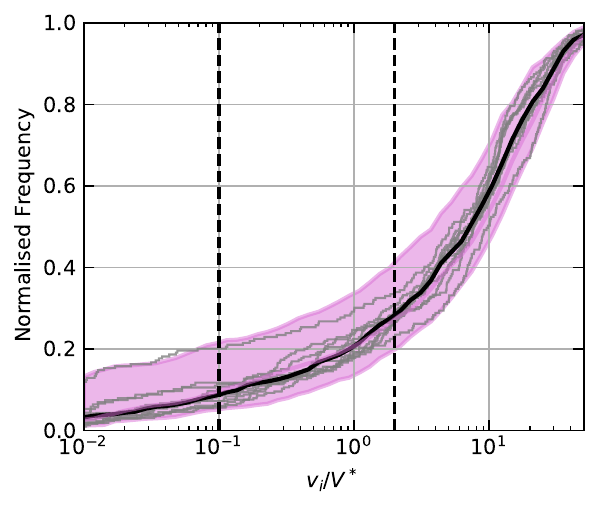}
    \caption{Same as Fig. \ref{fig:norm_freq_vel_ratio_no_rem_many} but now for disks 3 to 12 in Table \ref{tab:disk_table} which are formed from giant impact 1 at 0.5 au with an orientation of $0\pi$. The purple area is the $2\sigma$ range of the median fit to disks 13 to 52 which are formed the same giant impact but has orientation of $0.5\pi$. The figure compares collision outcomes between the 10 simulated disks for giant impact 1 with an orientation of $0\pi$ against an orientation of $0.5\pi$.}
    \label{fig:vel_ratio_0pi}
\end{figure}

Fig. \ref{fig:vel_ratio_0pi} shows the $v_i/V^*$ for disks 3 to 12 in Table \ref{tab:disk_table} in grey. The median distribution is shown in black, and the purple area shows the 2.5th to 97.5th percentile range for the $v_i/V^*$ in Fig. \ref{fig:norm_freq_vel_ratio_no_rem_many}. Between the different orientations, we find no significant differences between the number of destructive collisions between planetesimals. Like in the different sma cases, the disks formed from giant impact 1 with different orientations have the same absolute velocity distribution. The relative velocity difference between the planetesimal groups will only be down to random sampling the parent population of ejecta from the SPH simulation. There is a difference in the distribution of the directions the velocity kicks are given due to differing orientations but this has no bearing on the absolute velocity values.  
    
\begin{figure}
    \centering
    \includegraphics[width=\linewidth]{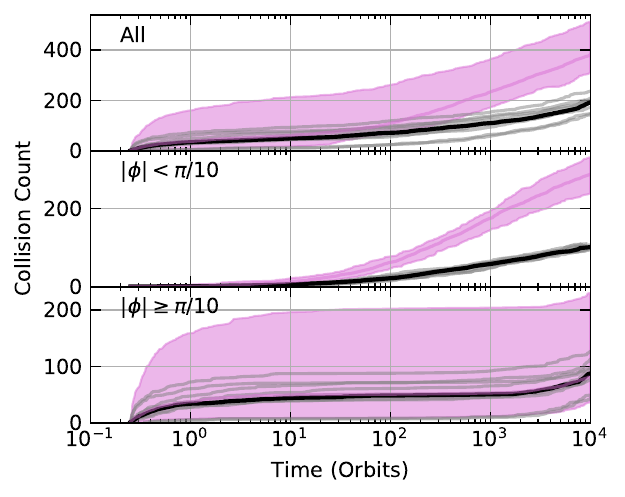}
    \caption{The cumulative collision count for disks 3 to 12 in table \ref{tab:disk_table} which are formed from giant impact 1 at 0.5 au with an orientation of $0\pi$ in grey. The median collision count is shown in black, and the pruple filled area represents the $2\sigma$ range about the median collision line for disks 13 to 52 which are formed the same giant impact but has orientation of $0.5\pi$. The figure is split into three parts, the top shows all destructive collisions between planetesimals, the middle shows collisions that happened within an azimuthal cut of $|\phi| <\pi/10$, and the bottom shows collisions that happened outside the collision point.}
    \label{fig:des_col_count_all_orientation}
\end{figure}

The orientation of the giant impact does change the number of collisions within the resulting disk. We show in Fig. \ref{fig:des_col_count_all_orientation} how the number of collisions in disks 3 to 12 varies over time compared to disks 13 to 52 which are shown in Fig. \ref{fig:des_col_count_all}. Between the two sets of simulations, we find that the median number of collisions in disks formed from giant impact 1 with an orientation of $0\pi$ to be approximately half the number of collisions that occurred in disks that formed from an orientation of $0.5\pi$. The difference is caused by the number of collisions at the collision point in the two sets of simulations. A decrease in the number of collisions at the collision point indicates that the number density in the $0\pi$ disk simulation set is reduced compared to the number density of the $0.5\pi$ simulation set. Fig \ref{fig:number_density_sma} shows the median number density of disks formed from giant impact 1 with orientations of $0\pi$ and $0.5\pi$, varying in impact sma from 0.3 to 2 au. The overall behaviour of how the density evolves is the same regardless of the initial orientation of the giant impact, with the density falling off as expected from equation (\ref{eqn:number_density}). We find that there is no significant difference for the collisions that occur outside the collision point. We see this trend appear in Fig. \ref{fig:disk_mass_au} and Fig. \ref{fig:disk_mass_au_split} where the difference in the mass produced between the different orientations is caused solely by the mass produced at the collision point. The difference between the number of collisions at the collision point will be down to the difference in orbital parameters given to the planetesimals. The orbital parameter difference is caused by the difference in the direction of the orbital kicks given to the planetesimals. In the $0\pi$ orientation, the kicks given to the planetesimals are dominated by radial velocity kicks. Radial velocity kicks are less efficient at changing the orbit of a particle than tangential velocity kicks. The $0.5\pi$ orientation has the velocity kicks mostly in the tangential direction, hence the orbital parameter difference and therefore different disk structure. For the $0\pi$ orientation the planetesimals are on more similar orbits when compared to $0.5\pi$ orientation, reducing the number of potential orbital crossings between planetesimals. 

\begin{figure}
    \centering
    \includegraphics[width=\linewidth]{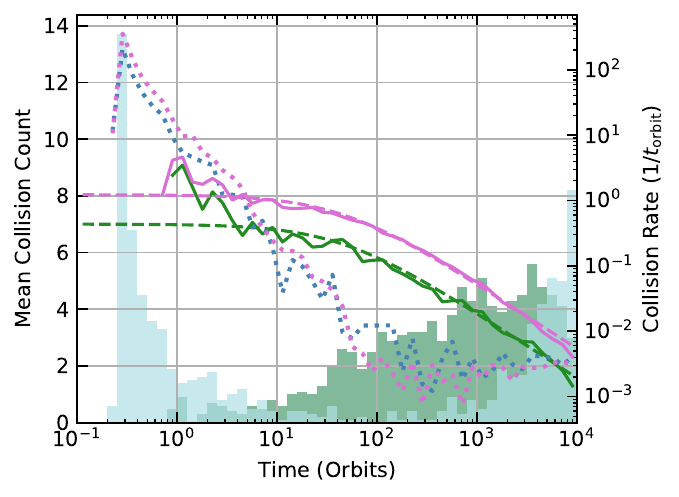}
    \caption{Left: The mean destructive collision count histograms for disks 3 to 12 which are formed from giant impact 1 at 0.5 au with an orientation of $0\pi$ in grey. Right: The mean destructive collision rate calculated from the bin count divided by the bin width, lines track collision rates. Green represents collisions occurring in an azimuthal cut of $|\phi|<\pi/10$ centred on the collision point, and blue represents collisions that occur outside the collision point. The purple represents mean destructive collision rate around the collision point for disks 13 to 52  which are formed the same giant impact but has orientation of $0.5\pi$. The solid lines follow the measured collision rate around the collision point, dotted lines track the collision rate outside the collision point, and the dashed lines represent the expected collision rate around the collision point from equation (\ref{eqn:des_colp_col_rate}).}
    \label{fig:col_rate_compare_rot}
\end{figure}

Fig. \ref{fig:col_rate_compare_rot} shows the mean collision rate around the collision point for the disks 3 to 12 (green) and disks 13 to 52 (purple) as solid lines. The mean collision rate outside the collision point is shown with blue and purple dotted lines for disks 3 to 12 and disks 13 to 52 respectively. The dashed lines show the expected collision rate using equation (\ref{eqn:des_colp_col_rate}). There is no difference between the evolution of the collision rates between the different orientations of giant impact 1, we only find a difference in the initial collision rate with $0.5\pi$ orientation starting with a larger collision rate around the collision point. Even within the first few orbits the collision rate is the same, suggesting the remnant is a large influence early on. Once the collision rate outside the collision point stops being the dominant collision position, the collision rate around the collision point settles to the expected collision rate. We find there is no difference in how a disk formed from different orientations of a giant impact evolves over time. The only difference between disks is the initial set up which sets the initial collision rate. This is seen in how the density at the collision point evolves over time in Fig. \ref{fig:number_density_sma}. The density at the collision point is typically lower for the $0\pi$ orientation over $10^4$ orbits, though the density change is not as drastic as when the sma is varied. 

\subsubsection{Varying Impact}

Giant impact 1 is not the only impact that can occur. We have conducted a preliminary test of the parameter space to look at two other impact scenarios. These impacts are listed in Table \ref{tab:giant_impacts}. In summary, impact 3 is the same as impact 1 with a higher impact velocity, while impact 2 is a more typical collision with a mass ratio of 0.4, and an impact parameter of $b=0.4$. 

\begin{figure}
    \centering
    \includegraphics[width=\linewidth]{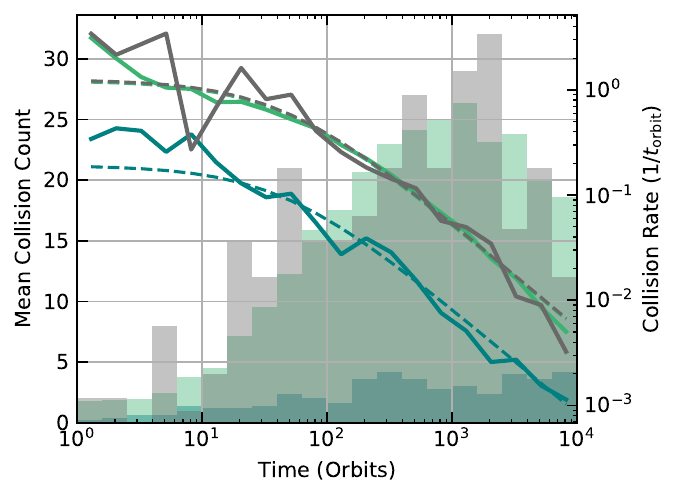}
    \caption{Same as Fig. \ref{fig:des_col_rate_au} but now shows the mean destructive collision count for the three giant impacts listed in Table \ref{tab:giant_impacts}. The disks used are listed as 13-52, 59-63, and 64 in Table \ref{tab:disk_table}. Giant impacts 1, 2, and 3 are green, blue and grey respectively. }
    \label{fig:colp_many_col}
\end{figure}

In Fig. \ref{fig:colp_many_col} we show histograms for the collision count and the collision rate for the three giant impacts with orientations of $0.5\pi$ and placed at 0.5 au. We find for giant impacts 1 and 3 that they have similar behaviour which is to be expected as the only varying parameter is the impact velocity between the two. We find giant impact 2 to be less collisionally active. At late times we see that all giant impacts follow the expected collision rate at the collision point. We see deviations early on due to small number statistics with giant impact 2 and the effect of the remnant in giant impacts 1 and 3. We note that in the giant impact 2 case, the second largest remnant had a large enough mass to be considered important to track gravitationally. The two large remnants in giant impact 2 might be the cause of the reduced collision rate. Another important difference with giant impact 2 is the change in distribution of the planetesimals initially. The difference in distributions is what is likely to cause the large difference in collision rates between the different disks. We need to explore more parameter space in order to definitively say how differing giant impact parameters would affect the collision rate in the disk. We have shown in these initial giant impacts that the collision rate in the disk can vary greatly even with planetesimal disks having similar masses (giant impact 2 produces a disk with 70\% of the mass of giant impact 1 and 3). We also show that the evolution of the collision rate at the collision point is consistent across different giant impacts.  

    \label{fig:num_dens_au_dist}

\section{Discussion}
\label{sec:discussion}

We now discuss how our results affect the observability of an EDD. We also discuss what the limitations of the study are and potential improvements. 

\subsection{Flux}

Now we know how a planetesimal disk formed after a giant impact might collisionally behave over time, we can look at the effect of the planetesimal disk on the observed extreme debris disk. It is complicated to model the effect of debris produced in planetesimal-planetesimal collisions on the observability and track how that debris evolves over time. There are many factors to consider such as how the planetesimals form solid objects post-giant impact. Are they a collection of small rocks forming rubble piles, or completely formed solid bodies? In the early disk lifetime, it could be possible that some of each planetesimal has not fully solidified or aggregated. Throughout we have assumed that the planetesimals are solid bodies which are the best producers of small grains as shocks are able to pass through them more easily. In the example of a rubble pile, the object is more likely to break up into smaller chunks but nothing that would be immediately visible. In any collision, we do not know what the size distribution of the ejecta will be. 

\begin{figure}
    \centering
    \includegraphics[width=\linewidth]{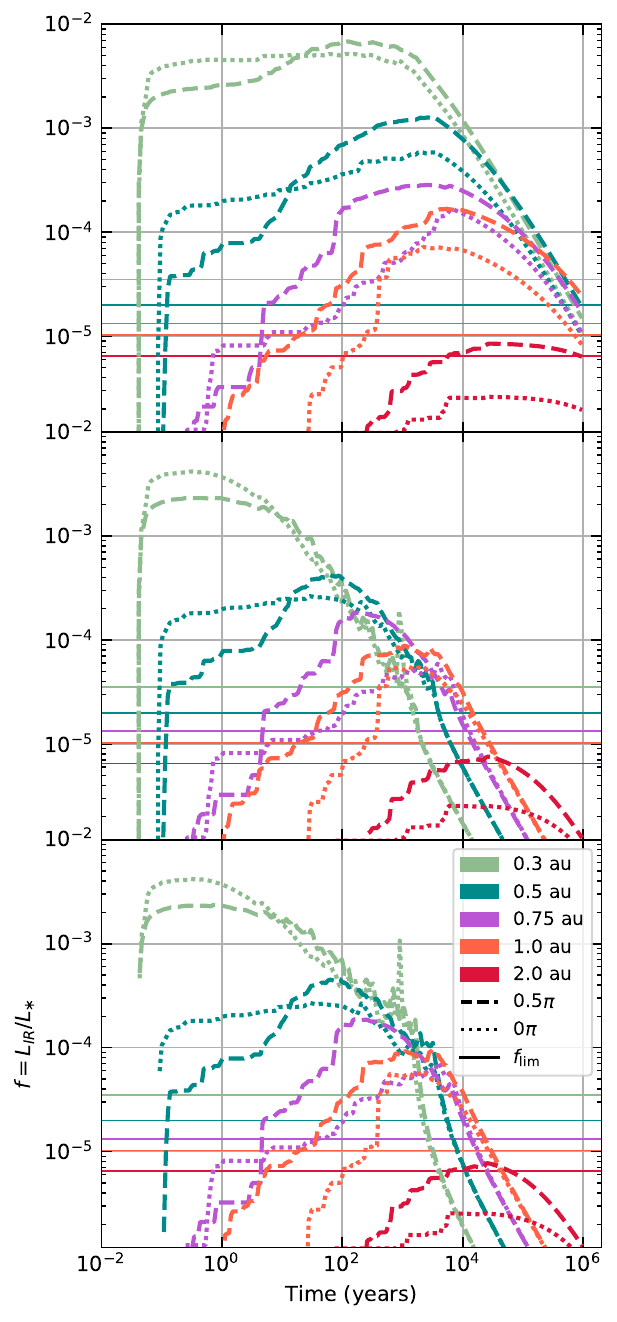}
    \caption{The estimated flux \citep{Wyatt-2008-evolution-of-debris-disks} of the planetesimal disk  using three alternative methods for giant impact 1 with a orientation of $0.5\pi$ (dashed) and $0.0\pi$ (dotted). The planetesimal disks are placed between 0.3 and 2 au. The solid horizontal lines dictate the detection limit dependent on disk position. 
     The disks all had the same initial velocity distribution and only differed in location and giant impact orientation. Disks represented in plot can be found in Table \ref{tab:disk_table} and are listed starting from the top downwards as follows: 2, 13, 54, 56, and 58 are the dashed lines, and 1, 3, 53, 55, and 57 are the dotted lines.
    Top: timescale is set by equation (\ref{eqn:tc_plan}) (method 1). Middle: timescale is set by equation (\ref{eqn:t_c_time_variable}) (method 2). Bottom: mass and timescale are set by equations \ref{eqn:mass_update_no_des}, \ref{eqn:mass_update_des}, and \ref{eqn:t_c_update_des} (method 3).}
    \label{fig:est_flux}
\end{figure}

To simplify we can estimate the fractional luminosity of debris produced by a planetesimal disk by using a traditional debris disk model. Fig \ref{fig:est_flux} shows a fractional luminosity over time plot for disks between 0.3 and 2 au formed from giant impact 1 with orientations $0.5\pi$ and $0\pi$ which are the dashed and dotted lines respectively.
The dashed and dotted lines are the smoothed expected fractional flux where the mass is added at the time each collision occurs. The fractional luminosity for each disk is calculated using the model from \citet{Wyatt-2008-evolution-of-debris-disks} that describes the evolution of a traditional debris disk,
\begin{equation}
    \label{eqn:fraclum}
    f/M_{\rm tot} = 0.37r^{-2}D_{bl}^{-0.5}D_c^{-0.5},
\end{equation}
where $f$ is the fractional luminosity of the disk, $M_{\rm tot}$ is the total mass in the disk in Earth masses, $r$ is the position of the disk in au, $D_c$ is the diameter of the largest planetesimal in km and $D_{bl}$ is the blowout size in \micron. 
We calculate the expected size of the largest object using equation (44) from \citet{Leinhardt-2012-collisions-between-gravity-dom-bodies} and the median $v_i/V^*$ of collisions in disks 13 to 52, giving an approximate value of $1\ \rm{km}$. We stress that this value is a rough estimate as no current scaling-law scales to the extreme $v_i/V^*$ values we find in our simulations. This scaling law does not take into account that material will potentially be vaporised and, therefore, might inhibit the size of the largest object. There is a good possibility that the largest object could have a size much smaller than $1 \rm{km}$. We are also using an average value for the largest-sized object, while in reality the largest-sized object in each planetesimal collision will differ. 
We set a blowout size for a solar-like star of $0.8\ \rm{\mu m}$. To evolve a debris disk, all values in equation (\ref{eqn:fraclum}) are fixed besides $M_{\rm tot}$ which varies with time as:
\begin{equation}
    \label{eqn:m_tot_fill}
        M_{\rm{tot}}(t) = M_{\rm{tot}}(0)/[1+(t-t_{\rm stir})/t_c],
\end{equation}
where,
\begin{equation}
    \label{eqn:tc_full}
    t_c = 1.4\times 10^{-9}r^{13/3}(dr/r)D_cQ_D^{*5/6}e^{-5/3}M_*^{-4/3}M_{\rm{tot}}^{-1},
\end{equation}
$dr/r$ is the width of the disk which is set to 0.5, $Q^*_D$ is the planetesimal strength assumed to be 150 $\rm{J\ kg^{-1}}$, $e$ is the mean planetesimal eccentricity of the planetesimal disk, and $M_*$, the central star mass, is one solar mass. The timescale of mass loss ($t_c$) starts when the destructive collisions occur which is determined by $t_{\rm stir}$. We set $t_{\rm stir}$ to be the time at which the first planetesimal collision occurs so that $t = 0 $ refers to the time at which the giant impact occurred.

The above describes a fully formed traditional debris disk and how it evolves over time.
To know how the fractional luminosity behaves after adding mass from planetesimal-planetesimal collisions, we need to estimate how the mass from each collision evolves. To do so we define method one by making equation (\ref{eqn:m_tot_fill}) a summation of each planetesimal collision:
\begin{equation}
    \label{eqn:m_tot_plan}
            M_{\rm{tot}}(t) =
           \sum_{i}^{N_{\rm tot}} m_i(0)/[1+(t-t_{\rm stir})/t_{c,i}],
\end{equation}
where,
\begin{equation}
    \label{eqn:tc_plan}
    t_{c,i} = 1.4\times 10^{-9}r^{13/3}(dr/r)D_cQ_D^{*5/6}e^{-5/3}M_*^{-4/3}m_i^{-1},
\end{equation}
here $m_i$ is the mass liberated from each destructive planetesimal-planetesimal collision, $i$, $N_{\rm tot }$ is the total number of destructive planetesimal-planetesimal collisions, and all other variables keep the same values as above. With equation (\ref{eqn:m_tot_plan}) and (\ref{eqn:tc_plan}) we can add mass at any arbitrary time. The issue that arises with our method here is we are evolving the mass from different planetesimal-planetesimal collisions separately. This is not realistic as the mass liberated from one collision will affect the evolution of the mass liberated from a different collision. We could not have added mass at any arbitrary time using equation (\ref{eqn:m_tot_fill}) and (\ref{eqn:tc_full}) as the total mass is determined by the initial total mass and $t_c$ which is inversely proportional to $M_{\rm tot}$. The disk would evolve too quickly in this scenario. In our scenario, the mass in the disk will evolve slower than expected though it does allow for complex fractional luminosity behaviour. The complex luminosity behaviour is realistic compared to an ever steeper curve as mass is added and near instantly will be observable. The changes in the fractional luminosity will be near instantaneous after every planetesimal-planetesimal collision. In both cases, we do not take into account how the mass added to the observable disk from planetesimal collisions interacts with other material that forms from the giant impact or the underlying disk.

We can also estimate how the total mass within the disk varies over time with a $t_c$ that varies with time. We define method two with a variable timescale, $t_c(t)$, as 
\begin{equation}
\label{eqn:t_c_time_variable}
        t_{c}(t) = 1.4\times 10^{-9}r^{13/3}(dr/r)D_cQ_D^{*5/6}e^{-5/3}M_*^{-4/3}M_{\rm des}(t)^{-1},
\end{equation}
where $M_{\rm des}(t)$ is the total mass added to the observable disk up to time $t$. Now if we use equation (\ref{eqn:t_c_time_variable}) in equation (\ref{eqn:m_tot_plan}) the change in mass would be more drastic than if we use equation (\ref{eqn:tc_plan}). We note though that $t_c(t)$ is dependent on the mass added from each destructive planetesimal collision hence $t_c(t)$ will only decrease with time. Using equation \ref{eqn:t_c_time_variable} will cause depletion of mass from the disk on timescales shorter than what would be expected. 

A third method of estimating the total mass in the observable disk is to update the total mass when a destructive collision occurs. With method 3, the mass evolves over time as 
\begin{equation}
\label{eqn:mass_update_no_des}
    M_{\rm tot}(t) = M_{\rm tot}(t_i)/[1+(t-t_{stir})/t_c(t_i)],
\end{equation}
where $t_i$ is the time at which the $i^{\rm th}$ destructive planetesimal collision occurred. Mass is added to the disk when $t = t_{i+1}$, hence 
\begin{equation}
\label{eqn:mass_update_des}
    M_{\rm tot}(t_{i+1}) = \Delta M_{\rm{des}}(t_{i+1}) + M_{\rm tot}(t_i)/[1+(t_{i+1}-t_{stir})/t_c(t_i)],
\end{equation}
where $\Delta M_{\rm{des}}(t_{i+1})$ is the mass added to the disk from the $i+1^{\rm th}$ destructive planetesimal collision. We then find that $t_c$ must evolve when mass is added to the disk as
\begin{equation}
\label{eqn:t_c_update_des}
    t_{c}(t_{i+1}) = 1.4\times 10^{-9}r^{13/3}(dr/r)D_cQ_D^{*5/6}e^{-5/3}M_*^{-4/3}M_{\rm tot}(t_{i+1})^{-1}.
\end{equation}
The evolution of mass in the disk should now be more realistic as it adjusts the timescale $t_c$ over time and is not just constant for each planetesimal as in method one or does not decrease as sharply as in method two. We therefore expect methods one and two to give us an approximate upper and lower bounds for the flux evolution with method 3 providing a more realistic case. 

In Fig. \ref{fig:est_flux} we show how the observed planetesimal disks varies with time for all three methods we have outlined. We find that the giant impact location has a massive effect on the flux of the extreme debris disk. The closer in the giant impact, the greater the effect of the planetesimal disk on the fractional luminosity of the extreme debris disk. We expected a brighter disk closer in as the disk overall will be hotter, but this is not the only reason. The planetesimal disk that will feed the extreme debris disk material is more collisionally active closer to the central star. Planetesimal disks closer to the star are likely to have a larger number of collisions happen earlier in the lifetime of the disk, leading to an initial large increase in dust added to the observable extreme debris disk. The vapour condensate disk should be observable near instantaneously after the giant impact \citep{Watt_2021_vapour_EDD}, and planetesimal disks close to the star will provide more material for the extreme debris disk. Hence, extreme debris disks formed from a giant impact close to their host star will have a larger fractional luminosity not just because the disk will be hotter but there will be more mass in small grains provided by the planetesimal population being collisionally active early in the lifetime of the disk. 

The difference we would expect to see from orientation of the giant impact does not make a real difference in terms of fractional luminosity for disks close to the star but has a larger affect the further the giant impact occurs from the star. We see in section \ref{sec:orientation} that there is no difference between the early collision rate of the disk when orientated differently. The difference only arises in the collision rate at the collision point. For giant impact 1, as it is placed closer to the star the collisional activity in the initial few orbits increases and produces a substantial amount of mass. The mass produced from collisions in the early disk allow the $0\pi$ orientations to have similar fractional luminosities to the $0.5\pi$ cases in disks placed at 0.3 and 0.5 au. Once the giant impact is placed at a distance which reduces the initial collision rate then the difference between $0\pi$ and $0.5\pi$ cases is established. We see there is a significant difference between the different orientations from the 0.75 au case outwards in Fig. \ref{fig:est_flux}. 

We find there to be a stark difference between the fractional luminosity evolution depending on whether we use method one with each destructive planetesimal collision being set a collisional timescale or method two with which all planetesimal mass follows a single collisional timescale that varies with time. The third method follows similar behaviour to the second method at early times and there is not much difference between the fractional luminosities but has a longer observable lifetime. 
With method one we find that the fractional luminosity tends to be brighter and remains in a brightened state for longer than compared to equation method two or three which we expected. Interesting to note that for method one we find that nearly all disks regardless of distance from the star will be observable for at least $\sim 10^5$ years after the giant impact if we assume a detection threshold $R_{\nu} = 0.03$ where $R_{\nu} = F_{\nu \ \rm{disk}}/F_{\nu *}$, where $F_{\nu \ \rm{disk}}$ and $F_{\nu *} $ are the flux from the debris disk and the star respectively. The detection limit then would be \citep{Wyatt-2008-evolution-of-debris-disks}
\begin{equation}
    f_{\rm det} = 6\times 10^9 R_{\nu}L_*T^{-4}_*B_{\nu}(\lambda,T_*)[B_{\nu}(\lambda,T)]^{-1}X_{\lambda}, 
\end{equation}
where r is the distance to the disc from the star, $L_*$ is the luminosity
of the star, $T_*$ is the blackbody temperature of the star, T is the
temperature of the disc, $B_{\lambda}$ is the blackbody emission, and $X_{\lambda}$ is a
factor included to take account of the falloff in the emission spectrum
at large wavelengths but it is 1 for 24 $\rm{\mu m}$. The detection limit for at $\lambda = 24\ \rm{\mu m}$ is used as the cutoff in fig. \ref{fig:est_flux}.
For method two the time at which the disk becomes undetectable varies with distance from the star, with the 0.3 au disks having the shortest detectable lifetimes. Method two gives us an estimate of the lower limit on the detectability lifetimes of the disks while equation method two is likely the upper limit on the detectability lifetime. Method three gives us a more realistic estimate of the disk lifetimes. It is worth keeping in mind that the $N$-body simulations did not extend above $10^4$ orbits of the progenitor. For disks closer to the star, this means a a longer period of mass being removed from the observable disk without mass being added from the planetesimal disk, which is why we see similar trends at late times between methods two and three. 


\subsection{Disk Mass}

The observable mass produced by collisions between planetesimals in our simulated disks will not account for all the observable mass produced in a real planetesimal disk. First, we do not account for the mass between our planetesimal size distribution ($>$10 km) and the vapour condensate ($<$100 mm). While the debris that forms with sizes between these two populations (we call intermediate debris) does not make up a large fraction of the total mass in the disk, it is a reservoir of material which could affect the flux of the disk. While initially it will not be observable, disks formed by giant impacts will have a greater collision rate enhanced by the collision point. The intermediate debris can be ground down on a quick timescale to be observable. From the SPH simulations, we define the intermediate debris as the total mass of particles with a vapour fraction between 10\% and 20\%. Here we assume the vapour fraction is significant enough to hinder the growth of larger melt objects so this material will form smaller melt objects. 

The intermediate debris was not looked at as it will not provide a substantial amount of material to the visible disk over a large amount of time. Though this is dependent on the assumed vapour fraction cuts. The disks formed from giant impacts are asymmetric, it is difficult to exactly know when this mass will be added to the observable disk. Compared to a traditional debris disk, the collision point in a giant impact induced disk increases the collisional activity within the disk. Since we know the vapour condensates would be seen near instantaneously after the giant impact and planetesimals add to the disk flux after a few orbits we would expect the intermediate debris to be processed relatively quickly, therefore adding to the flux of the disk on the order of a few orbits. Since the top grain/boulder size (biggest being metres in size) of the intermediate debris will be small, once it is collisionally active the lifetime will be short so we can assume all the mass will be added at once. If the fraction of intermediate debris is comparable to the vapour mass then it will have a significant impact on the fractional luminosity seen from the extreme debris disk. More work needs to be done on the assumption that giant impacts will form different populations of vapour condensates/intermediate debris/planetesimals and how these populations interact with each other. 

\subsection{Mass definition}

We have defined three different populations formed after a giant impact: a vapour condensate disk seen near instantaneously after the impact, a planetesimal population which is formed by near purely melt material, and the intermediate debris formed from a mixture of melt and vapourised material. We made the assumption that for any object made up of more than 10\% vapour material will not be able to clump into a planetesimal-like object as the expanding vapour bubble(s) will impede this behaviour. However, this is not a known value hence how does the behaviour of the planetesimal disk change with the vapour cutoff value? Changing the vapour cutoff value will change the mass used to draw planetesimals from in our simulations. A larger vapour cutoff will increase the planetesimal mass and vice versa. From the SPH simulations of the giant impacts we are able to define the largest planetesimals in our simulations and from these planetesimals we extend the size distribution downwards until we have distributed the total escaping particle mass below the vapour cutoff into planetesimals. By changing the vapour cutoff value, we change the the smallest planetesimal size in our distribution as we fix the size distribution to be $dN \propto D^{-3}$. Changing the vapour cutoff value changes the number of planetesimals at the smaller end of the planetesimal distribution, as the smaller planetesimals are sampled from a size distribution with an upper size set by the smallest grouped planetesimal. The collisional activity in a simulated disk is related to the number of objects in that disk. Therefore, changing the vapour cutoff value will change the likelihood of collisions between planetesimals occurring, not just the mass in the disk overall. In our scenario with a fixed size distribution, a larger vapour cutoff will lead to a disk with more mass that is collisionally more active. 

If we fix the number of planetesimals when changing the vapour cutoff fraction, therefore changing the size distribution, this will also affect the collisional activity in the disk as it will increase the size of the non-grouped planetesimals.
The larger the planetesimal is, the more likely a collision will occur with it. An increase in mass due to a increase in the vapour cutoff value will also lead to a more collisionally active disk as there will be an increase in larger planetesimals. Though the collisional outcomes will differ depending on what we fix. Fixing the size distribution will lead to a change in the number of non-grouped planetesimals which are more easily disrupted at lower impact velocities (at least for planetesimals in the gravity regime). Fixing the number of planetesimals will increase the size of non-grouped planetesimals which take greater impact velocities to disrupt. In reality, neither will be fixed and the size distribution of the planetesimals formed from a giant impact will most likely not be a smooth power law. As the size distribution is unknown, we decide to fix it so that the planetesimal disk can be compared to traditional debris disk behaviour. 

\subsection{Gravity}
In our $N-$body simulations of planetesimal disks we have not allowed gravitational interactions between planetesimals to take place. It is a simplification which allowed a much greater compute speed. No gravity between planetesimals would reduce the collision rate within the whole disk. Gravity would act to enlarge the cross-section of each planetesimal due to gravitational focusing, therefore the collisional cross-section of each planetesimal would be larger than the physical size. So with gravity it is like each planetesimal having an inflated radius, with the larger planetesimals having their radii inflated more than the smaller planetesimals. \citet{Jackson-2014-planetary-collisions-at-large-au} has an expression for the collision rate of a single planetesimal as $R_{\rm{col}} = n\sigma v_{\rm{rel}}$ where $n$ is the number density, $\sigma$ is the cross-section and $v_{\rm{rel}}$ is the velocities of other planetesimals relative to the planetesimal we are measuring the collision rate for.  The collision rate varies with a factor of $\sigma$, hence we should not expect to find a difference in how the collision rate changes with time. We would only expect there to be an increased collision rate initially before decreasing as in equation (\ref{eqn:des_colp_col_rate}), so no change in the turning point, $t_p$, but we would expect to see a change in the initial collision rate, $a$.

\subsection{Impact Angle}

We have assumed that collisions between planetesimals are all head-on collisions. In reality this would not be the case. The average impact angle between two bodies is $45^{\circ}$. The omission of other impact angles was because we could not accurately determine the impact angle between two bodies in our $N$-body simulations. Our simulations then became a test case for the most extreme disk we could have as head-on impacts require lower impact velocities to disrupt. If we accounted for impact angles then we would expect the value of $V^*$ to increase in many of the recorded collisions. The increase in $V^*$ would shift some destructive impacts down into the bouncing regime, and some in the bouncing regime into the merging regime. The rate of collisions would not change but the outcome of individual collisions might. Also the median collision velocity is much greater than the limit of $2V^*$ that even with the addition of impact angle we would expect most collisions to still be destructive. We would expect the observed disk formed of material produced from destructive planetesimals to be reduced if we include non head-on collisions. The observed disk would potentially peak at a lower fractional luminosity as a consequence. Though this is likely not going to be a significant difference.

\subsection{The Full Disk}
It was proposed by \citet{Su-2019-extreme-disk-variability} that the flux behaviour seen in ID8 after 2017 could be caused by a planetesimal population from giant impacts which could have occurred at an earlier time in the lightcurve in 2013 and 2014. We discussed the possibility in \citet{Watt_2021_vapour_EDD} of a planetesimal population and how the fractional luminosity might evolve using the \citet{Wyatt-2008-evolution-of-debris-disks} model for debris disks. We assumed that a planetesimal population would take some time to collisionally grind enough material to produce debris to add to the disk. We made a conservative estimate that the process of creating small debris from planetesimals would take 100 orbits. 

In this paper, we have shown that planetesimals can contribute to a debris disk on much shorter timescales. We have found that the planetesimal disk formed from a giant impact is highly destructive with most collisions exceeding our criterion for full destruction of both planetesimals in a disk. The collision rate is tied to the sma which the giant impact occurred at. For a giant impact that occurs between 0.3 and 0.5 au we would expect a highly active planetesimal disk producing a substantial amount of observable dust grains. ID8 has suspected impacts between 0.3 and 0.5 au and therefore the planetesimal disk should be highly active. It is fully possible that the impacts which could have occurred in 2013 and 2014 produced a planetesimal disk which was feeding material to the EDD in 2017. 

We have not explored how the vapour condensate disks which we studied in \citet{Watt_2021_vapour_EDD} and the planetesimal disks will interact. It is possible that the dust added to the EDD from the planetesimals might reduce or wipe out any periodic behaviour in the EDD. It may also enhance the periodic behaviour with planetesimals overwhelmingly colliding at the collision point. Further study is needed on how these two disks interact. 

\section{Conclusions}
\label{sec:conclusion}
The aim of this paper was to better understand how the ejecta from a giant impact formed and evolved EDDs. We focused on the formation of planetesimals post-giant impact from the ejecta and how the mass passed from the planetesimals into small grains which are observable. The idea to study planetesimal activity was spurred on by the behaviour of the lightcurve from ID8 which has a period of increasing excess flux that is not attributed to an impact in 2017. We used a mixed simulation that involved modelling giant impacts using SPH to calculate the ejected mass and distribution. We then assumed that the planetesimals formed from the melted ejecta. We used $N-$body ({\sc rebound}) to evolve the planetesimal distribution spatially and collisionally. 

From the $N-$body simulations we have shown that planetesimal disks are collisionally active at early times in the disk lifetime. We studied the parameter space varying sma, the giant impact used to form disks, and the orientation of the giant impact. The mass produced in a disk can be different depending on the randomly sampled distribution given to the planetesimal population. Though behaviour and evolution of the planetesimal disk is consistent across all parameters. We provide equations that describe the number of collisions, collision rate and number density at the collision point. The collision point is the dominant location for collisions over the $10^4$ orbits for which we simulated each disk. There is variation in the number of collisions outside the collision point which can affect the mass produced and the collision rate outside the collision point varies with sma. The collision rate outside the collision point is initially linked to the number density around the remnant(s) within the disk. For giant impacts closer to the star, planetesimals need larger kicks in order to escape the influence of the remnants(s). Hence at early times, disks that sit close to the star will have enhanced collision rates not expected from the collision rate given for the collision point. We find that the orientation of a giant impact does affect the collision rate at the collision point, but not as much as varying sma. 

We find that the collisionally active disks can have a significant influence on the fractional luminosity of the disk. For disks that sit closer to the star, the larger collision rate produces more mass early on that can pass down to the observable disk. Hence, not only are the disks warmer but there is more mass that can be observable. We show that a planetesimal disk can sustain an EDD on a timescale before the vapour condensate material is removed from the disk. Further study is needed on how the vapour condensate and the dust from the planetesimal collision interact. Especially there is a need to understand how the periodic behaviour of some EDDs is affected by mass being added from a planetesimal population within the disk. 

\section*{Acknowledgements}

 LW acknowledges financial support from STFC (grant S100048-102). ZML and PJC acknowledge financial support from the Science and Technology Facilities Council (grant number: ST/V000454/1). This work was carried out using the computational facilities of the Advanced Computing Research Centre, University of Bristol - http://www.bris.ac.uk/acrc/. 

\section*{Data Availability}

The data is available on request to the author.



\bibliographystyle{mnras}
\bibliography{planet_bib} 



\appendix

\section{Table of Gadget2 embryo setup values}

\begin{table*}
    \label{tab:gadget2setup}
    \caption{Summary of the set up values used to equilibrate planetary embryos used in the giant impact SPH simulations. GI Index indicates what embryo was used in giant impacts listed in Table \ref{tab:giant_impacts}; $N_{\rm{tot}}$ is the number of SPH particles in the embryo; Mass is the mass of the embryo in $10^{-1}$ Earth masses; Radius is the radius of the embryo in $10^{-1}$ Earth radii; $S_{\rm{core}}$ is the specific entropy set for the core entropy cooling in $10^7$ ergs $\rm{g}^{-1}\ \rm{K}^{-1}$; $S_{\rm{mantle}}$ is the specific entropy set for the mantle entropy cooling in $10^7$ ergs $\rm{g}^{-1}\ \rm{K}^{-1}$; $T_{\rm{forced}}$ is the equilibration time with specific entropy values and velocity dampening is forced in $10^4$s; $T_{\rm{equil}}$ is the equilibration time without any forced values in $10^4$s.}
    \begin{tabular}{ c c c c c c c c c }
        \hline
        Index & GI Index & $N_{\rm{tot}}$ & Mass & Radius & $S_{\rm{core}}$ & $S_{\rm{mantle}}$ & $T_{\rm{forced}}$ & $T_{\rm{equil}}$ \\
        $-$ & $-$ & $-$ & $10^{-1}\ \rm{M_{\oplus}}$ & $10^{-1} \rm{R_{\oplus}}$ & $10^7$ ergs $\rm{g}^{-1}\ /rm{K}^{-1}$ & $10^7$ ergs $\rm{g}^{-1}\ \rm{K}^{-1}$ & $10^4$s & $10^4$s\\
        \hline \hline
        1 & 1 & 200000 & 1.185 & 5.242 & 1.58 & 2.24 & 7.2 & 7.2 \\
        2 & 2 & 285715 & 2.525 & 6.648 & 1.61 & 2.30 & 5.0 & 2.6 \\
        3 & 2 & 114285 & 1.185 & 5.214 & 1.58 & 2.24 & 3.0 & 2.0 \\
        4 & 3 & 20000 & 0.995 & 4.879 & 1.36 & 2.74 & 7.2 & 7.2 \\
        \hline 
    \end{tabular}
\end{table*}

\section{Table of Disk Simulations}

\begin{table*}
\caption{Summary of planetesimal disk set up and results from {\sc rebound} simulations. GI Index is the giant impact reference to the giant impact used to set up the disk in table \ref{tab:giant_impacts}; $N_{\rm{tot}}-$ total number of planetesimals in the simulation;  $M_{\rm{des}}-$ the total mass produced in the simulation through destructive planetesimal-planetesimal collisions; sma - the semi-major axis the giant impact occurred at to set up the disk; rot - the orientation of the giant impact when setting up the disk; $N_{\rm{p-p}}-$ the number of planetesimal-planetesimal collisions; \%des - the percentage of destructive collisions between planetesimals; \%bounce - the percentage of bouncing collisions between planetesimals; \%merge - the percentage of merging collisions between planetesimals; $N_{\rm{lr-p}}-$ the number of collisions between the remnant(s) and planetesimals.
\newline
$^*$ Disks with the same initial velocity distribution of planetesimals before orientation of the giant impact is taken into account.}
\label{tab:disk_table}
\begin{tabular}{ c c c c c c c c c c c }
\hline
Index & GI Index & $N_{\rm{tot}}$ & $M_{\rm{des}}$ & sma & rot & $N_{\rm{p-p}}$ & \%des & \%bounce & \%merge & $N_{\rm{lr-p}}$ \\
$-$ & $-$ & $-$ & ($10^{-5}\rm{M_{\oplus}}$) & (au) & $\pi$ & $-$ & $-$ & $-$ & $-$ & $-$\\
\hline\hline
1$^*$ & 1 & 19845 & 360.73 & 0.30 & 0.00 & 818 & 78.24 & 15.77 & 5.99 & 533\\
\hline 
2$^*$ & 1 & 19845 & 383.87 & 0.30 & 0.50 & 1371 & 65.21 & 10.14 & 24.65 & 725\\
\hline 
3$^*$ & 1 & 19845 & 79.83 & 0.50 & 0.00 & 262 & 74.05 & 19.85 & 6.11 & 308\\
4 & 1 & 19845 & 56.01 & 0.50 & 0.00 & 187 & 79.14 & 9.09 & 11.76 & 261\\
5 & 1 & 19845 & 80.92 & 0.50 & 0.00 & 255 & 78.82 & 13.33 & 7.84 & 243\\
6 & 1 & 19845 & 76.18 & 0.50 & 0.00 & 198 & 78.28 & 16.16 & 5.56 & 230\\
7 & 1 & 19845 & 74.86 & 0.50 & 0.00 & 259 & 73.36 & 18.92 & 7.72 & 313\\
8 & 1 & 19845 & 77.10 & 0.50 & 0.00 & 209 & 68.42 & 11.96 & 19.62 & 285\\
9 & 1 & 19845 & 61.73 & 0.50 & 0.00 & 259 & 76.45 & 16.22 & 7.34 & 251\\
10 & 1 & 19845 & 118.98 & 0.50 & 0.00 & 302 & 78.15 & 16.56 & 5.30 & 303\\
11 & 1 & 19845 & 76.07 & 0.50 & 0.00 & 261 & 75.48 & 17.62 & 6.90 & 219\\
12 & 1 & 19845 & 93.00 & 0.50 & 0.00 & 227 & 74.45 & 14.98 & 10.57 & 245\\
\hline 
13$^*$ & 1 & 19845 & 152.36 & 0.50 & 0.50 & 412 & 75.97 & 11.65 & 12.38 & 245\\
14 & 1 & 19845 & 176.33 & 0.50 & 0.50 & 512 & 77.15 & 13.87 & 8.98 & 382\\
15 & 1 & 19845 & 298.34 & 0.50 & 0.50 & 422 & 81.99 & 11.85 & 6.16 & 404\\
16 & 1 & 19845 & 195.60 & 0.50 & 0.50 & 431 & 72.16 & 16.71 & 11.14 & 345\\
17 & 1 & 19845 & 206.13 & 0.50 & 0.50 & 517 & 72.53 & 13.35 & 14.12 & 350\\
18 & 1 & 19845 & 200.93 & 0.50 & 0.50 & 469 & 82.94 & 12.37 & 4.69 & 356\\
19 & 1 & 19845 & 158.00 & 0.50 & 0.50 & 396 & 80.05 & 8.59 & 11.36 & 351\\
20 & 1 & 19845 & 249.36 & 0.50 & 0.50 & 521 & 78.12 & 12.86 & 9.02 & 448\\
21 & 1 & 19845 & 209.90 & 0.50 & 0.50 & 335 & 75.82 & 19.40 & 4.78 & 99\\
22 & 1 & 19845 & 218.94 & 0.50 & 0.50 & 581 & 67.30 & 14.80 & 17.90 & 241\\
23 & 1 & 19845 & 183.65 & 0.50 & 0.50 & 483 & 72.46 & 20.70 & 6.83 & 331\\
24 & 1 & 19845 & 211.63 & 0.50 & 0.50 & 488 & 77.25 & 12.91 & 9.84 & 383\\
25 & 1 & 19845 & 148.49 & 0.50 & 0.50 & 424 & 79.48 & 11.56 & 8.96 & 302\\
26 & 1 & 19845 & 148.27 & 0.50 & 0.50 & 465 & 79.35 & 13.98 & 6.67 & 416\\
27 & 1 & 19845 & 215.02 & 0.50 & 0.50 & 452 & 83.41 & 11.28 & 5.31 & 328\\
28 & 1 & 19845 & 144.07 & 0.50 & 0.50 & 523 & 71.89 & 15.11 & 13.00 & 388\\
29 & 1 & 19845 & 163.25 & 0.50 & 0.50 & 525 & 77.71 & 15.43 & 6.86 & 373\\
30 & 1 & 19845 & 205.85 & 0.50 & 0.50 & 563 & 73.53 & 19.89 & 6.57 & 312\\
31 & 1 & 19845 & 182.74 & 0.50 & 0.50 & 481 & 82.33 & 11.02 & 6.65 & 380\\
32 & 1 & 19845 & 172.12 & 0.50 & 0.50 & 733 & 68.76 & 23.74 & 7.50 & 402\\
33 & 1 & 19845 & 186.81 & 0.50 & 0.50 & 636 & 67.45 & 15.57 & 16.98 & 362\\
34 & 1 & 19845 & 207.42 & 0.50 & 0.50 & 453 & 77.04 & 12.14 & 10.82 & 308\\
35 & 1 & 19845 & 155.72 & 0.50 & 0.50 & 457 & 75.27 & 11.82 & 12.91 & 310\\
36 & 1 & 19845 & 137.74 & 0.50 & 0.50 & 479 & 80.38 & 15.24 & 4.38 & 374\\
37 & 1 & 19845 & 166.40 & 0.50 & 0.50 & 476 & 84.03 & 12.39 & 3.57 & 346\\
38 & 1 & 19845 & 207.93 & 0.50 & 0.50 & 576 & 74.13 & 17.19 & 8.68 & 373\\
39 & 1 & 19845 & 312.64 & 0.50 & 0.50 & 1335 & 62.77 & 26.82 & 10.41 & 370\\
40 & 1 & 19845 & 140.96 & 0.50 & 0.50 & 420 & 77.62 & 14.29 & 8.10 & 477\\
41 & 1 & 19845 & 167.36 & 0.50 & 0.50 & 464 & 77.16 & 13.36 & 9.48 & 292\\
42 & 1 & 19845 & 215.60 & 0.50 & 0.50 & 763 & 64.35 & 14.29 & 21.36 & 286\\
43 & 1 & 19845 & 200.43 & 0.50 & 0.50 & 525 & 77.71 & 14.86 & 7.43 & 316\\
44 & 1 & 19845 & 119.73 & 0.50 & 0.50 & 434 & 73.27 & 12.44 & 14.29 & 354\\
45 & 1 & 19845 & 166.46 & 0.50 & 0.50 & 532 & 73.87 & 19.36 & 6.77 & 406\\
46 & 1 & 19845 & 138.18 & 0.50 & 0.50 & 447 & 76.73 & 15.66 & 7.61 & 377\\
47 & 1 & 19845 & 225.87 & 0.50 & 0.50 & 735 & 62.99 & 17.01 & 20.00 & 334\\
48 & 1 & 19845 & 137.91 & 0.50 & 0.50 & 433 & 76.44 & 16.17 & 7.39 & 404\\
49 & 1 & 19845 & 212.88 & 0.50 & 0.50 & 521 & 71.79 & 20.92 & 7.29 & 332\\
50 & 1 & 19845 & 156.07 & 0.50 & 0.50 & 413 & 80.63 & 14.04 & 5.33 & 353\\
51 & 1 & 19845 & 174.44 & 0.50 & 0.50 & 579 & 77.72 & 15.20 & 7.08 & 356\\
52 & 1 & 19845 & 201.52 & 0.50 & 0.50 & 605 & 77.02 & 17.69 & 5.29 & 373\\
\hline 

\end{tabular}
\end{table*}

\begin{table*}
    \contcaption{}
    \label{tab:continued}
    \begin{tabular}{ c c c c c c c c c c c }
\hline
Index & GI Index & $N_{\rm{tot}}$ & $M_{\rm{des}}$ & sma & rot & $N_{\rm{p-p}}$ & \%des & \%bounce & \%merge & $N_{\rm{lr-p}}$ \\
$-$ & $-$ & $-$ & ($10^{-5}\rm{M_{\oplus}}$) & (au) & ($\pi$) & $-$ & $-$ & $-$ & $-$ & $-$\\
\hline
53$^*$ & 1 & 19845 & 30.04 & 0.75 & 0.00 & 102 & 83.33 & 8.82 & 7.84 & 138\\
\hline 
54$^*$  & 1 & 19845 & 64.25 & 0.75 & 0.50 & 180 & 79.44 & 11.67 & 8.89 & 75\\
\hline 
55$^*$  & 1 & 19845 & 22.85 & 1.00 & 0.00 & 50 & 80.00 & 8.00 & 12.00 & 93\\
\hline 
56$^*$  & 1 & 19845 & 57.72 & 1.00 & 0.50 & 144 & 84.72 & 11.11 & 4.17 & 137\\
\hline 
57$^*$  & 1 & 19845 & 2.56 & 2.00 & 0.00 & 12 & 75.00 & 16.67 & 8.33 & 21\\
\hline 
58$^*$  & 1 & 19845 & 8.40 & 2.00 & 0.50 & 34 & 82.35 & 11.76 & 5.88 & 55\\
\hline 
59 & 2 & 12559 & 34.34 & 0.50 & 0.50 & 86 & 72.09 & 18.60 & 9.30 & 58\\
60 & 2 & 12559 & 33.61 & 0.50 & 0.50 & 79 & 75.95 & 11.39 & 12.66 & 55\\
61 & 2 & 12559 & 40.90 & 0.50 & 0.50 & 87 & 66.67 & 22.99 & 10.34 & 86\\
62 & 2 & 12559 & 42.37 & 0.50 & 0.50 & 76 & 75.00 & 14.47 & 10.53 & 68\\
63 & 2 & 12559 & 52.90 & 0.50 & 0.50 & 112 & 68.75 & 16.07 & 15.18 & 55\\
\hline 

64 & 3 & 13545 & 209.94 & 0.50 & 0.50 & 418 & 77.75 & 11.72 & 10.53 & 193\\
\hline 
\hline \\ 
\end{tabular}
\end{table*}


\bsp	
\label{lastpage}
\end{document}